\documentclass[aps,pre,longbibliography,footinbib,twocolumn,pacs,preprintnumbers,notitlepage,amsmath,amssymb]{revtex4-1}

\usepackage{amssymb}
\usepackage{amsmath}

\usepackage{color}
\usepackage[pdftex]{graphicx}

\usepackage{empheq}

\usepackage{mathtools}

\usepackage{makecell}

\usepackage{enumerate}
\usepackage{mathtools}
\usepackage{stmaryrd}

\usepackage{enumitem}

\usepackage{textcomp}
\usepackage{gensymb}

\newcommand{\moy}[1]{\left\langle #1 \right\rangle}

\newcommand{\ex}[1]{\mathrm{e}^{#1}}

\newcommand{\dd}[0]{\mathrm{d}}
\newcommand{\ii}[0]{\mathrm{i}}
\newcommand{\kk}[0]{\boldsymbol{k}}

\newcommand{\rr}[0]{\boldsymbol{r}}

\newcommand{\xx}[0]{\boldsymbol{x}}

\newcommand{\yy}[0]{\boldsymbol{y}}
\newcommand{\pp}[0]{\boldsymbol{p}}

\newcommand{\kB}[0]{k_{\mathrm{B}}}

\newcommand{\kapx}[0]{\kappa_0}
\newcommand{\kapphi}[1]{\kappa_{#1}}

\definecolor{darkblue}{rgb}{0,0,0.6}
\definecolor{darkred}{rgb}{0.6,0,0}
\usepackage[colorlinks=true,urlcolor=darkblue,citecolor=darkblue,linkcolor=darkred]{hyperref}

\begin{document}

\title{Diffusion of a tracer in a dense mixture of soft particles \\connected to different thermostats}

\author{Marie Jardat}

\author{Vincent Dahirel}
%\affiliation{Sorbonne Universit\'e, CNRS, Physico-Chimie des \'Electrolytes et Nanosyst\`emes Interfaciaux (PHENIX), 4 Place Jussieu, 75005 Paris, France}

\author{Pierre Illien}
%\affiliation{Sorbonne Universit\'e, CNRS, Physico-Chimie des \'Electrolytes et Nanosyst\`emes Interfaciaux (PHENIX), 4 Place Jussieu, 75005 Paris, France}

\affiliation{Sorbonne Universit\'e, CNRS, Laboratoire PHENIX (Physico-Chimie des \'Electrolytes et Nanosyst\`emes Interfaciaux (PHENIX)), 4 Place Jussieu, 75005 Paris, France}

\date{\today}

\begin{abstract}

 We study the dynamics of a tracer in a dense mixture of particles connected to different thermostats. Starting from the overdamped Langevin equations that describe the evolution of the system, we derive the expression of the self-diffusion coefficient of a tagged particle in the suspension, in the limit of soft interactions between the particles. Our derivation, which relies on the linearization of the Dean-Kawasaki equations obeyed by the density fields and on a path-integral representation of the dynamics of the tracer,  extends previous derivations that held for tracers in contact with a single bath.  Our analytical result is confronted to results from Brownian dynamics simulations. The agreement with numerical simulations is very good even for high densities. We show how the diffusivity of tracers can be affected by the activity of a dense environment of soft particles that may represent polymer coils -- a result that could be of relevance in the interpretation of measurements of diffusivity in biological media. Finally, our analytical result is general and can be applied to the diffusion of tracers coupled to different types of fluctuating environments, provided that their evolution equations are linear and that the coupling between the tracer and the bath is weak.

\end{abstract}

\maketitle

%\tableofcontents

\section{Introduction}

Describing suspensions of interacting active particles (agents which are able to take up energy from their environment and to convert it into directed motion) has been a central challenge of statistical physics during the past decades, and has resulted in the design of different successful theoretical frameworks \cite{Vicsek2012,Marchetti2013, {Cates2015}, Bechinger2016}. More recently, going beyond the situation where all the particles in the suspension are identical, the question of mixtures of particles with different levels of activity has drawn a lot of attention.  Indeed, in various situations of physical or biological interest, one encounters situations where particles that are active, in the sense that they are very far from equilibrium, interact with `passive' particles, which are only submitted to the equilibrium thermal fluctuations of their environment. This is for example the case in the intracellular medium, where many different agents (organelles, proteins, enzymes...) have different levels of activity, and such heterogeneities are known to have a significant impact on the structure and dynamics of the cytoplasm \cite{Guo2014,Parry2014}.

From a theoretical perspective, a natural way to model these mixtures is to assume that the different groups of particles are in contact with different thermostats -- the simpler situation is that of a binary mixture of `hot' and `cold' particles. This concept has progressively attracted more and more attention in nonequilibrium statistical physics, and was explored numerically in colloidal suspensions \cite{{Weber2016},Tanaka2016a}, polymeric systems \cite{Smrek2018,Smrek2017,Chubak2020,Smrek2020}, {and in the context of the thermal Casimir effect \cite{Dean}}. From an analytical perspective, phase separation in mixtures of `hot' and `cold' particles was studied in the low-density limit, in which the system reduces to a two-body problem  \cite{Grosberg2015,Ilker2020}. The three-body problem for particles in contact with different thermostats was solved recently for specific pairwise interactions \cite{Wang2020}.

Although a lot of knowledge has been gathered about collective properties in mixtures of particles in contact with multiple thermostats,  little is known about the properties of tagged particles in such suspensions, in spite of their importance. For instance, the self-diffusion coefficient of a tracer is a key observable to describe the transport properties inside these complex systems, and may be of interest to interpret observations from experimental cell biology \cite{Guo2014,Parry2014}. So far, the long-time self-diffusion coefficient in mixtures of particles with different temperatures has only been investigated in the low-density limit, and in the case of short-range repulsive interactions between the particles \cite{Ilker2021}.

Here, we consider the general situation of a tracer whose diffusion is affected by its coupling to multiple fluctuating fields, in contact with different thermostats. In this setting, we derive the effective diffusion coefficient of the tracer in the small coupling limit, using a path-integral representation that was previously designed to study the dynamics of a tracer in contact with a single bath \cite{Demery2011}. In particular, we apply this formalism to the situation of a dense suspension of particles interacting via soft potentials, that we choose to be Gaussian soft-core potentials, which are relevant to describe the interactions between polymer coils \cite{Louis2000,Lang2000,Likos2001,Wensink2008}. We argue that this potential is also well adapted to model the diffusion of large tracers (organelles, macromolecules...) in the intracellular medium. Technically, the equations obeyed by the density fields are obtained using Ito calculation and adapting the usual Dean-Kasawaki derivation \cite{Kawasaki1994,Dean1996}, and are then linearized and solved for -- a technique that was used in different contexts over the past years (microrheology of colloidal suspensions \cite{Demery2014, Demery2019,Demery2015}, active matter  \cite{Feng2021,Poncet2021b,Tociu2019,Fodor2020,{Tociu2020}, Martin2018}, binary mixtures \cite{Poncet2016}, electrolytes \cite{Mahdisoltani2021a,Mahdisoltani2021,Demery2015a,Frusawa2020,Frusawa2022,Avni2022}).

Comparing with results from numerical simulations of the microscopic dynamics of the system, we check the validity of the approximations on which our analytical result relies. In the range of parameters investigated here, analytical results are always very close to numerical results, with a discrepancy that never exceeds $5\%$. We also show that the dynamics of tracers is significantly enhanced when they are placed in a `hot' bath. This effect, which relies on local energy transfer from the hotter to the colder particles, was evidenced numerically \cite{{Weber2016},Tanaka2016a}, and was described analytically in the low-density limit and for hard, repulsive, short-ranged potentials \cite{Ilker2021}. The present work therefore provides an analytical basis for this effect in the opposite limit of very soft particles, and for potentially very high densities -- two important aspects for the applicability of such theories in biological context. Finally, we emphasize that this formalism is very general, and can be used to describe diffusion in different kinds of fluctuating environments, such as membranes, colloidal suspensions, or more generally Gaussian fields with various prescriptions for the relevant order parameters that can either be non-conserved  or conserved (`model A' and `model B' dynamics, respectively \cite{Chaikin,Hohenberg1977}).

\section{Model}

\begin{figure}
\begin{center}
\includegraphics[width=0.7\columnwidth]{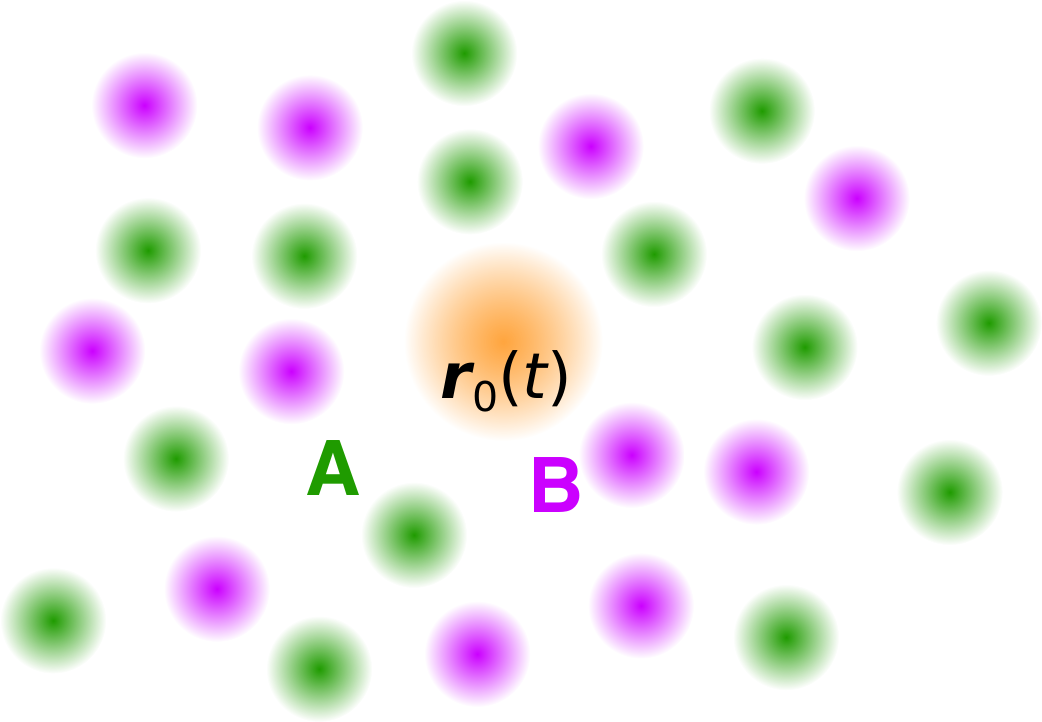}\\
\vspace{1cm}
\includegraphics[width=0.7\columnwidth]{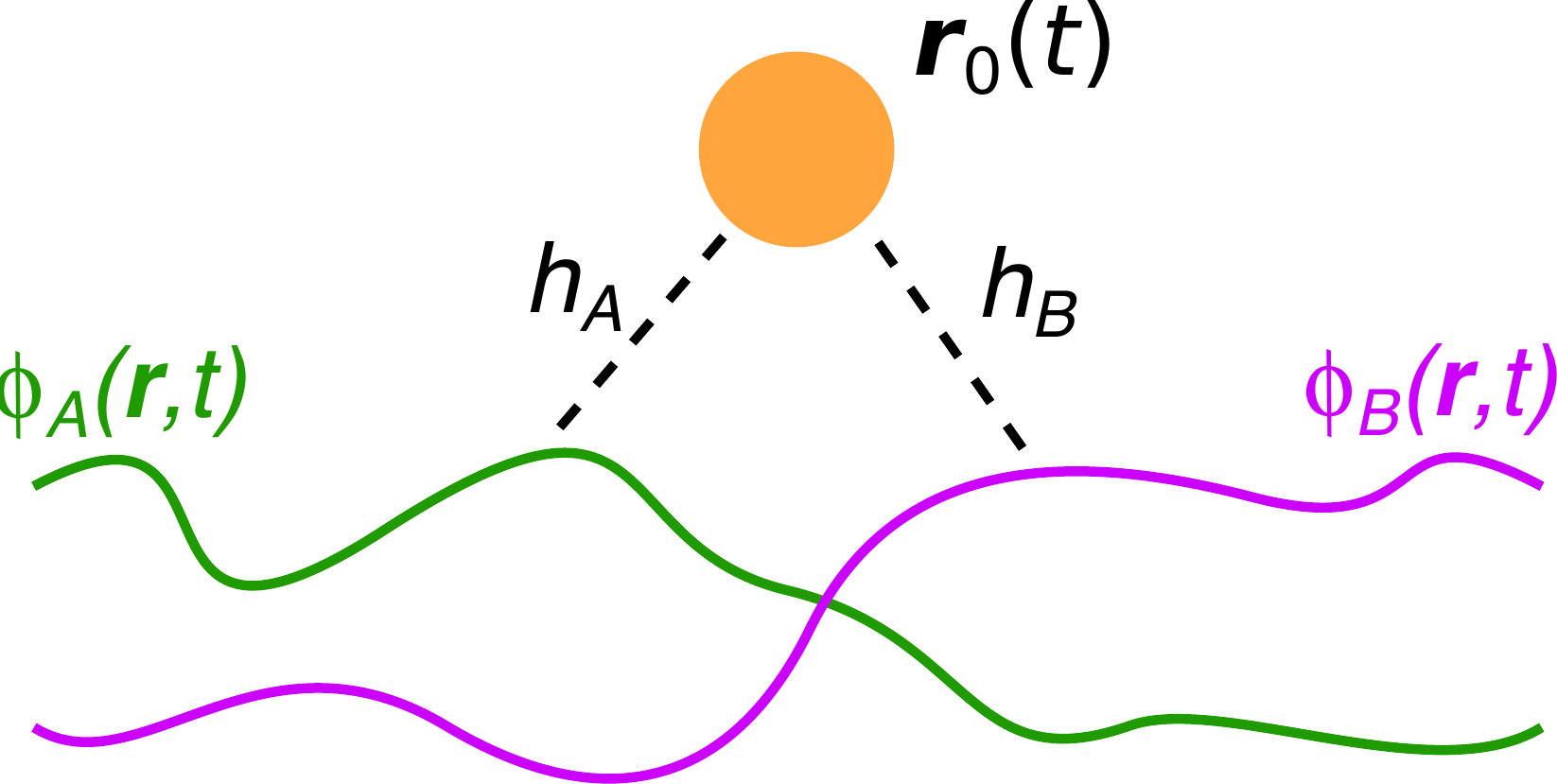}
\caption{Top: System under study and considered in the main text: a tracer, whose position at time $t$ is denoted by $\rr_0(t)$ is coupled to a binary mixture constituted of particles of type A and B. Bottom: General situation considered in Appendix \ref{app_Deff}: a tracer (at position $\rr_0(t)$) is coupled to two fluctuating fields $\phi_A(\rr,t)$ and $\phi_B(\rr,t)$. The parameters $h_A$ and $h_B$ quantify the intensities of the couplings.}
\label{fig_system}
\end{center}
\end{figure}

We consider a tracer, whose position is denoted by $\rr_0(t)$, and which interacts with a bath of $N$ particles, whose positions are denoted by $\rr_1(t),\dots,\rr_N(t)$ (Fig. \ref{fig_system}). The bath particles can be of different types. We assume that there are $\mathcal{N}$ different types, and that $N_\alpha$ denotes the number of particles of type $\alpha$, in such a way that $N = \sum_{\alpha=1}^\mathcal{N} N_\alpha$. We assume that each particle in the system obeys an overdamped Langevin dynamics, and that the evolution of the system is given by the $N+1$ coupled equations:
\begin{equation}
\label{overdampedLangevin}
\frac{\dd \rr_a}{\dd t} = -\kappa_\alpha \sum_{b =0}^N \nabla V_{\alpha \beta }(\rr_a-\rr_b) + \sqrt{2 D_\alpha}\boldsymbol{\zeta}_a(t),
\end{equation}
where $ V_{\alpha \beta}$ denotes the pair interaction potential between two particles $a$ of type $\alpha $ and $b$ of type $\beta$ (we will use the notation   $ V_{\alpha 0}$ to denote the interaction between the tracer and a particle of type $\alpha$), and to simplify the notation we use the convention $\nabla V_{\alpha \beta} (0)=0$. The bare diffusion coefficient of a particle of type $\alpha$ is related to the mobility $\kappa_\alpha$ through the Einstein relation $D_\alpha = \kB T_\alpha \kappa_\alpha$. Note that the mobility of the particles is assumed to be independent of the temperature, in such a way that the bare diffusion coefficient of each species is proportional to the temperature of the corresponding thermostat. The noise terms $\boldsymbol{\zeta}_a(t)$ have the following properties:
\begin{align}
\langle \zeta_{a,i}(t) \rangle &= 0,\\
\langle \zeta_{a,i}(t)\zeta_{b,j}(t') \rangle &= \delta_{ab} \delta_{ij} \delta(t-t').
\end{align}

In order to coarse-grain the dynamics, we define the density of bath particles of type $\alpha$ as 
\begin{equation}
\rho_\alpha(\xx,t) = \sum_{\substack{\text{particles $a$}\\\text{of type $\alpha$}}}\delta (\rr_a(t)-\xx).
\end{equation}
Using Ito calculation \cite{Gardiner1985}, and relying on the usual derivation proposed by Dean for a single-component fluid \cite{Dean1996} {and later extended for binary mixtures \cite{Demery2015a,Poncet2016}}, we obtain the coupled equations for the fields $\rho_\alpha$
\begin{align}
&\partial_t \rho_\alpha    =\sqrt{2D_\alpha} \nabla \cdot [\boldsymbol{\eta}_\alpha \sqrt{\rho_\alpha}] + D_\alpha \nabla^2 \rho_\alpha    \nonumber \\
&+\kappa_\alpha \nabla \cdot \left[   \rho_\alpha  \sum_{\beta=1}^\mathcal{N}\nabla(V_{\alpha\beta}\ast \rho_\beta) +\rho_\alpha \nabla(V_{\alpha 0}\ast \delta_{\rr_0}) \right], \label{Dean_A}
\end{align}
with
\begin{equation}
\label{ }
\langle \eta_{\alpha,i} (\xx,t)\eta_{\beta,j}(\xx',t')\rangle = \delta_{\alpha\beta} \delta_{ij} \delta(\xx-\xx') \delta(t-t').
\end{equation}
The symbol $\ast$ represents spatial convolution:
\begin{equation}
\label{ }
(f\ast g)(\xx) = \int \dd \yy \; f(\yy) g(\xx-\yy),
\end{equation}
 and we use the shorthand notation $\delta_{\rr_0}(\xx)=\delta(\xx-\rr_0)$.   The evolution of the tracer position is given by the  equation of motion:
\begin{equation}
\label{ }
 \frac{\dd}{\dd t}\rr_0(t) = -\kappa_0 \sum_{\alpha=1}^\mathcal{N}\nabla(V_{\alpha 0}\ast\rho_\alpha)(\rr_0(t),t)+ \sqrt{2D_0}\boldsymbol{\xi}(t)
\end{equation}
with the noise $\langle \xi_i(t)\xi_j(t')\rangle=\delta_{ij}\delta(t-t')$, and where $\kappa_0$ is the bare mobility of the tracer. The goal of the calculation is to determine the mean-square displacement of the tracer and its self-diffusion coefficient in the long-time limit, defined as
\begin{equation}
\label{definition_Deff}
D_\text{eff} = \lim_{t\to\infty} \frac{\langle [\rr_0(t)-\rr_0(0)]^2 \rangle}{2d t},
\end{equation}
 where $d$ is the spatial dimension {(see Section \ref{sec:convergence} in Appendix \ref{app_Deff} for a detailed discussion on the conditions required for normal diffusion to be observed in this model)}. From now on, we consider the case of a binary mixture ($\mathcal{N}=2$). Eq. \eqref{Dean_A} is written for $\alpha = A,B$ and is linearized, defining $\phi_i$ as $\rho_i = \bar{\rho}_i+\sqrt{\bar{\rho}_i}\phi_i$. This technique was introduced to study tracer diffusion in colloidal suspensions \cite{Demery2014}, and allows one to retrieve results from the `random phase approximation' \cite{Likos2001,Lang2000,Louis2000} when applied to compute static quantities, such as pair correlation functions.
 
  We define the total density ${\rho}= \bar{\rho}_A+\bar{\rho}_B$ and the fraction $X$ such that $\bar{\rho}_A=X{\rho}$ and $\bar{\rho}_B = (1-X){\rho}$. At leading order in $\phi_i$, we find 
\begin{align}
\label{}
&\partial_t \phi_A =    \sqrt{2D_A} \nabla \cdot \boldsymbol{\eta}_A +D_A \nabla^2 \phi_A \nonumber \\
 &+\kappa_A \left[  \nabla^2 (X v_{AA}\ast \phi_A)+\nabla^2 (\sqrt{X(1-X)} v_{AB}\ast \phi_B) \right. \nonumber\\
&\left.  + \frac{\sqrt{\bar{\rho}_A}}{{\rho}} \nabla^2 ( v_{A0}\ast \delta_{\rr_0})   \right] 
\end{align}
(and similarly for $\phi_B$) where we defined $v_{\alpha\beta} = {\rho} V_{\alpha\beta}$. We adopt the following conventions for Fourier transformation:
\begin{align}
\label{}
  \tilde{f}(\kk)  & = \int \dd \xx \; \ex{-\ii \kk \cdot \xx } f(\xx),   \\
{f}(\xx)  & = \int \frac{ \dd \kk}{(2\pi)^d} \; \ex{\ii \kk \cdot \xx }   \tilde{f}(\kk) .
\end{align}
In Fourier space, the coupled equations for $\tilde{\phi}_A(\kk,t)$ and $\tilde{\phi}_B(\kk,t)$ then read
\begin{align}
\label{sol_phi_Fourier}
& \partial_t 
 \begin{pmatrix}
     \tilde{\phi}_A(\kk,t)     \\
     \tilde{\phi}_B(\kk,t)   
\end{pmatrix}   = -\boldsymbol{m} \begin{pmatrix}
     \tilde{\phi}_A(\kk,t)     \\
     \tilde{\phi}_B(\kk,t)   
\end{pmatrix} 
  \nonumber \\
    &-k^2
    \begin{pmatrix}
      \sqrt{\frac{X}{\bar{\rho}}} {\ex{-\ii \kk \cdot \rr_0(t)}}\kappa_A\tilde{v}_{A0}    \\
         \sqrt{\frac{1-X}{\bar{\rho}}} {\ex{-\ii \kk \cdot \rr_0(t)}}\kappa_B \tilde{v}_{B0}
\end{pmatrix}  
+
\begin{pmatrix}
     \sqrt{2D_A}\ii \boldsymbol{k}\cdot \tilde{\boldsymbol{\eta}}_A     \\
      \sqrt{2D_B} \ii \boldsymbol{k}\cdot  \tilde{\boldsymbol{\eta}}_B
\end{pmatrix},
\end{align}
with
\begin{equation}
\boldsymbol{m}= k^2 
\begin{pmatrix}
{\kB T_A}\kappa_A+  \kappa_A X \tilde{v}_{AA}    &   \kappa_A\sqrt{X(1-X)} \tilde{v}_{AB}    \\
   \kappa_B \sqrt{X(1-X)} \tilde{v}_{AB}  &   {\kB T_B}\kappa_B +  \kappa_B(1-X) \tilde{v}_{BB} 
\end{pmatrix}.
\label{m_explicite}
\end{equation}
After linearisation, the equation for the position of the tracer reads
\begin{align}
\label{eq_tracer_lin}
& \frac{\dd}{\dd t}\rr_0(t) = - \kappa_0 \sqrt{\frac{X}{\bar{\rho}}} \nabla(v_{A0}\ast\phi_A)(\rr_0(t),t) \nonumber\\
&- \kappa_0 \sqrt{\frac{1-X}{\bar{\rho}}} \nabla(v_{B0}\ast\phi_B)(\rr_0(t),t) +\sqrt{2D_0}\boldsymbol{\xi}(t)
\end{align}
To summarize, we show through Eq.  \eqref{eq_tracer_lin} how the dynamics of the tracer is linearly coupled to the density fields associated to each type of particle that constitute the bath of particles. These density fields obey a linear set of equations, which is written explicitly in Fourier space [Eqs. \eqref{sol_phi_Fourier}].

\section{Effective diffusion coefficient}

Although we managed to find a simple equation of motion for the tracer, which couples its position and the density fields associated with bath particles, computing its mean-square displacement is still a complicated task. Indeed, the position $\rr_0$(t)  of the tracer, which obeys Eq. \eqref{eq_tracer_lin} actually affects the evolution of the density fields, whose dynamics depend explicitly on $\rr_0$ through Eqs. \eqref{sol_phi_Fourier}. Treating this non-trivial coupling between the dynamics of the tracer and that of the field can be achieved in the small-coupling limit. We rely on the calculation that was done by Dean and D\'emery in the situation where a tracer is coupled to a single field \cite{Demery2011}, and extend it to the present situation, where the tracer is coupled to a binary mixture. {The derivation of the effective diffusion coefficient of the tracer in arbitrary dimension and in the limit of weak coupling is given in Appendix \ref{app_Deff}. In three dimensions, the result reads}
\begin{widetext}
\begin{equation}
{%\color{blue}
\frac{ \overline{D}_\text{eff} }{D_0}=1-\sum_{\alpha,\beta,\gamma} {\kappa_0 \kappa_\beta}  \int_0^\infty {\dd k}
\frac{ k^6}{6\pi^2}  \frac{\sqrt{X_\alpha X_\gamma}}{{\rho}} \widetilde{v}_{\alpha 0} \widetilde{v}_{\gamma 0}
\sum_{\nu=\pm1}   \frac{2 c^{(\nu)}_{\alpha\beta}}{(D_0 k^2 + \mu_\nu)^2}   \left[   \delta_{\gamma\beta} + \frac{T_\beta}{T_0}(D_0 k^2 - \mu_\nu) \sum_{\epsilon=\pm1}  \frac{c^{(\epsilon)}_{\gamma\beta}}{\mu_\nu + \mu_\epsilon}   \right]     ,
}
 \label{main_result_Deff}
\end{equation}
\end{widetext}
where we used the fluctuation-dissipation relation: $D_0= \kB T_0 \kappa_0$ and the fact that the integrand only depends on the modulus of $\kk$ to perform angular integrals. In this expression, $X_\alpha=X$ if $\alpha=A$ and $X_\alpha=1-X$ if $\alpha=B$. The eigenvalues $\mu_\pm$ are explicitly related to the physical parameters through the relation
\begin{align}
\label{eigenvalues_mu}
\mu_\pm =& \frac{k^2}{2} \left\{ \kappa_A(\kB T_A+X \tilde{v}_{AA} ) \right. \nonumber\\
&\left. + \kappa_B(\kB T_B+(1-X) \tilde{v}_{BB})\right\} \pm   \frac{s}{2}   ,
\end{align}
with
\begin{align}
\label{ }
s \equiv& k^2 \left\{\left[\kappa_A(\kB T_A+X \tilde{v}_{AA} )-\kappa_B(\kB T_B+(1-X) \tilde{v}_{BB} )\right]^2  \right. \nonumber\\
&\left. + 4 \kappa_A \kappa_BX(1-X)\tilde{v}_{AB}^2 \right\}^{1/2}.
\end{align}
The coefficients $c^{(\pm 1)}_{\alpha\beta}$ are the elements of the matrices
\begin{equation}
%\label{c_matrices}
\boldsymbol{c^{(\pm)} } =
\frac{1}{2s}
\begin{pmatrix}
{\pm m_{AA}\mp m_{BB}+s}
 &\pm 2{m_{AB}} \\
\pm 2{m_{BA}}
   & \mp m_{AA}\pm m_{BB}+s
\end{pmatrix},
\end{equation}
where the matrix $\boldsymbol{m}$ was defined in Eq. \eqref{m_explicite}.

Eq. \eqref{main_result_Deff} is the central result of the present work. Several comments follow:  (i) This expression is explicit in terms of all the parameters of the problem (interaction potentials between the different species, mobility coefficients, temperatures of the thermostats...) and can then be evaluated easily by performing the integral numerically; (ii) It was derived using a very general scheme, in such a way that its general expression (see Appendix \ref{app_Deff} and in particular Eqs. \eqref{Deff_compact} and  \eqref{Dalphabeta}) is applicable to other situations and may describe the diffusion of a tracer coupled to different fields, provided that the Hamiltonian of the system is quadratic in the fields $\phi_\alpha$ and that the tracer-field couplings are linear; {(iii) The convergence of the integrands in Eq.  \eqref{main_result_Deff} (and therefore the existence of normal diffusion) actually depends on the small-$k$ behavior of the rescaled potentials $\widetilde{v}_{\alpha\beta}(k)$. This is discussed in Section \ref{sec:convergence}.}

We emphasize that the $A-B$ mixture is stable as long as both eigenvalues $\mu_\pm$ [Eq. \eqref{eigenvalues_mu}] stay positive, to ensure that the solutions of Eq. \eqref{sol_phi_Fourier} do not diverge. $\mu_+$ is always positive, and the condition $\mu_- \geq 0$ reads
\begin{equation}
\label{stability_condition}
(\kB T_A+X \tilde{v}_{AA} )  (\kB T_B+(1-X) \tilde{v}_{BB} )  \geq  X(1-X) \tilde{v}_{AB}^2.
\end{equation}
In the specific case where $T_A=T_B=T$, the stability condition simplifies to
\begin{equation}
\label{stability_condition}
 (1+X\widetilde{u}_{AA})(1+(1-X)\widetilde{u}_{BB}) \geq X(1-X) \widetilde{u}_{AB}^2 ,
\end{equation}
where we define $ \widetilde{u}_{\alpha\beta} =  \widetilde{v}_{\alpha\beta}/(\kB T)$. In all the situations considered below, we choose parameters where the mixture remains stable.

\section{Some limit cases}

We now consider a few asymptotic limits of the general expression of the effective diffusion coefficient in three dimensions [Eq. \eqref{main_result_Deff}]. {We assume here that the interaction potentials are such that all the $k$-integrals written in this Section converge, which implies that the diffusion of the tracer is normal. An example of such a potential will be given and studied in details in the next Section.}

\subsection{Low-density limit}

We first consider the low-density limit ($\rho\to0$) of Eq. \eqref{main_result_Deff}, in which the result takes a simple form. In this limit, it is straightforward to show that $c^{(1)}_{\alpha\beta} = \delta_{\alpha,1}\delta_{\beta,1} + \mathcal{O}(\rho^2)$ and $c^{(-1)}_{\alpha\beta} = \delta_{\alpha,2}\delta_{\beta,2} + \mathcal{O}(\rho^2)$. Moreover, one gets from the definition of $\mu_{\pm}$ the following expansions: $\mu_+ = D_A k^2+  \mathcal{O}(\rho^2)$ and $\mu_- = D_B k^2+  \mathcal{O}(\rho^2)$. From Eq. \eqref{main_result_Deff}, this yields the following expression for the effective diffusion coefficient:
\begin{equation}
\frac{ \overline{D}_\text{eff} }{D_0}=1-\overline{\rho}\sum_{\alpha,\beta} \overline{D}^{(0)}_{\alpha\beta} + \mathcal{O}(\rho^2),
\end{equation}
with 
\begin{equation}
\overline{D}^{(0)}_{\alpha\beta} =  \delta_{\alpha\beta} \int_0^\infty  \dd k \frac{k^2 X_\beta \kappa_0 \widetilde{V}_{\beta,T}(\kk)^2(D_0-D_\beta+2\kB T_0 \kappa_\beta)}{6\pi ^2 \kB T_0(D_\beta+D_0)^2}.
\end{equation}
Interestingly, we observe that the correction to the diffusion coefficient does not involve the cross-terms $D_{AB}$ and $D_{BA}$, which only appear at order $\rho^2$.

\subsection{A tracer in contact with a  single hotter bath}

We then consider the particular situation where the tracer is a particle much larger than the bath particles, which are of a single type $A$, and which are connected to a hotter thermostat than the tracer ($T_A \gg T_0$). For simplicity, we can assume that the mobility of the particles are given by the Stokes-Einstein relation for a spherical particle $\kappa_\alpha = 1/(6\pi\eta \sigma_\alpha)$, where $\sigma_\alpha$ is the radius of the particle. Considering this particular case in Eq. \eqref{main_result_Deff} and taking the limit $T_A \gg T_0$ yields
\begin{equation}
\frac{D_\text{eff}}{D_0} = 1+\frac{\rho}{6\pi^2} \frac{\sigma_A}{\sigma_0} \int_0^\infty \dd k \; k^2 \frac{\tilde V_{0A}(k)^2}{(\kB T_0)(\kB T_A)}.
\end{equation}
The interaction potential $V_{0A}(r)$ is typically a function of the variable $r/(\sigma_0+\sigma_A)$, in such a way that its Fourier transform can be assumed to have the following form: $\tilde V_{0A} (k) = (\sigma_0+\sigma_A)^3\epsilon_{0A} f(k (\sigma_0+\sigma_A))$, where $\epsilon_{0A}$ is the typical interaction energy between the tracer and the bath particles, and $f$ is dimensionless. This yields 
\begin{equation}
\frac{D_\text{eff}}{D_0} = 1+\frac{\rho}{6\pi^2} \frac{\sigma_A(\sigma_0+\sigma_A)^3}{\sigma_0} \mathcal{A},
\end{equation}
where $\mathcal{A}$ is a dimensionless constant. It is interesting to deduce from this expression the typical root mean-square displacement of the tracer, rescaled by its bare value.  In the limit $\sigma_0 \gg \sigma_A$, one gets
\begin{equation}
\frac{\sqrt{D_\text{eff}}-\sqrt{D_0}}{\sqrt{D_0}} \sim \sigma_0.
\end{equation}
This scaling, which was observed experimentally for large tracers dispersed inside the cytoplasm of living cells \cite{Parry2014}, was also derived in the limit of a low density of crowders with purely repulsive  interactions \cite{Ilker2021}. Interestingly, it then appears that this scaling is robust against changes of the microscopic details of the model.

In the particular situation where all the particles have the same size and interact via the same potential $V$, and for an arbitrary temperature difference between the thermostat of the tracer and that of the bath, we get the following expression
\begin{align}
&\frac{D_\text{eff}}{D_0} = 1-\frac{\rho}{6\pi^2}  \nonumber\\
&\times\int_0^\infty \dd k \; k^2 \frac{\rho \tilde V (k)^2 [(2\theta-1)\rho \tilde V(k)^2 +(3\theta-1) \kB T_A]}{\theta [\rho \tilde V(k) + (\theta+1)\kB T_A][\rho \tilde V(k) + \kB T_A]},
\end{align}
where we introduced the ratio between the tracer and bath temperatures $\theta= T_0/T_A$. In the case where $\theta=1$, we retrieve previous results that were obtained in the equilibrium case where all the particles are connected to the same thermostat \cite{Demery2014}. We then consider the limit where the bath is much `hotter' than the tracer $\theta\ll1$: 
\begin{equation}
\frac{D_\text{eff}}{D_0} \underset{\theta\ll1}{=} 1+\frac{\rho}{6\pi^2 \theta} \int_0^\infty \dd k \; k^2 \tilde V(k)^2  +\mathcal{O}(1).
\end{equation}
In this situation, as expected intuitively, we find that ${D_\text{eff}}>{D_0}$: in other words, the hot bath enhances the diffusion of the tracer with respect to its bare value. In the opposite limit of $\theta\gg1$, we get
\begin{equation}
\frac{D_\text{eff}}{D_0} \underset{\theta\gg1}{=} 1-\frac{\rho}{2\pi^2 \theta^2} \int_0^\infty \dd k \; k^2 \tilde V(k)^2  +\mathcal{O}\left(  \frac{1}{\theta^3} \right),
\end{equation}
where, on the contrary, the diffusion of the tracer is hindered by the colder bath.

\section{Comparison with numerical simulations}
\label{sec_numerical_sim}

In order to go beyond the asymptotic analysis of limit cases, we now confront our analytical result to numerical simulations. We consider a binary mixture of Gaussian particles, which interact via the following potential
\begin{equation}
\label{Gaussiancore_def}
V_{\alpha\beta}(\rr) = \varepsilon_{\alpha\beta} \ex{-r^2/\sigma_{\alpha\beta}^2},
\end{equation}
and its Fourier transform:
\begin{equation}
\widetilde{V}_{\alpha\beta}(\kk)=\pi^{3/2}\sigma_{\alpha\beta}^3\varepsilon_{\alpha\beta}\, \ex{-k^2 \sigma_{\alpha\beta}^2/4}.
\end{equation}
This potential was introduced in the 1970s as a toy model to study phase transitions in suspensions of repelling particles \cite{Louis2000}, and its validity to describe polymer coils was discussed more recently \cite{Louis2000a}. The properties of the Gaussian-core fluid have been thoroughly studied through numerical simulations and approximate analytical approaches \cite{Louis2000,Lang2000,Likos2001,Wensink2008}, which makes it a good candidate to probe our analytical theory. {Note that the parameters $\varepsilon_{\alpha\beta}$ would generally be functions of the temperature, especially when these potentials are used to model polymer coils, but we assume here for simplicity that they do not depend on temperature.} {Finally, we emphasize that the $k$-dependence of this potential ensures the convergence of all the integration over Fourier modes, and therefore the existence of a normal diffusion regime.}

 In all the simulations presented below, {we consider three-dimensional systems,} we set $\sigma_{AA} = \sigma_{BB} = \sigma_{AB} = 1$, i.e. we assume that all particles have the same size, and that they have the same mobility $\kappa_{A} = \kappa_{B}  =\kappa$. This sets the unit length in our simulations.
 The evolution of the system is simulated using Brownian dynamics, which is a direct resolution of the coupled overdamped Langevin equations [Eq. \eqref{overdampedLangevin}] using the Euler scheme (see Appendix \ref{app_numerical} for details on the numerical simulations).

\subsection{Tracer in a single component fluid}

\begin{figure}
\begin{center}
\includegraphics[width=\columnwidth]{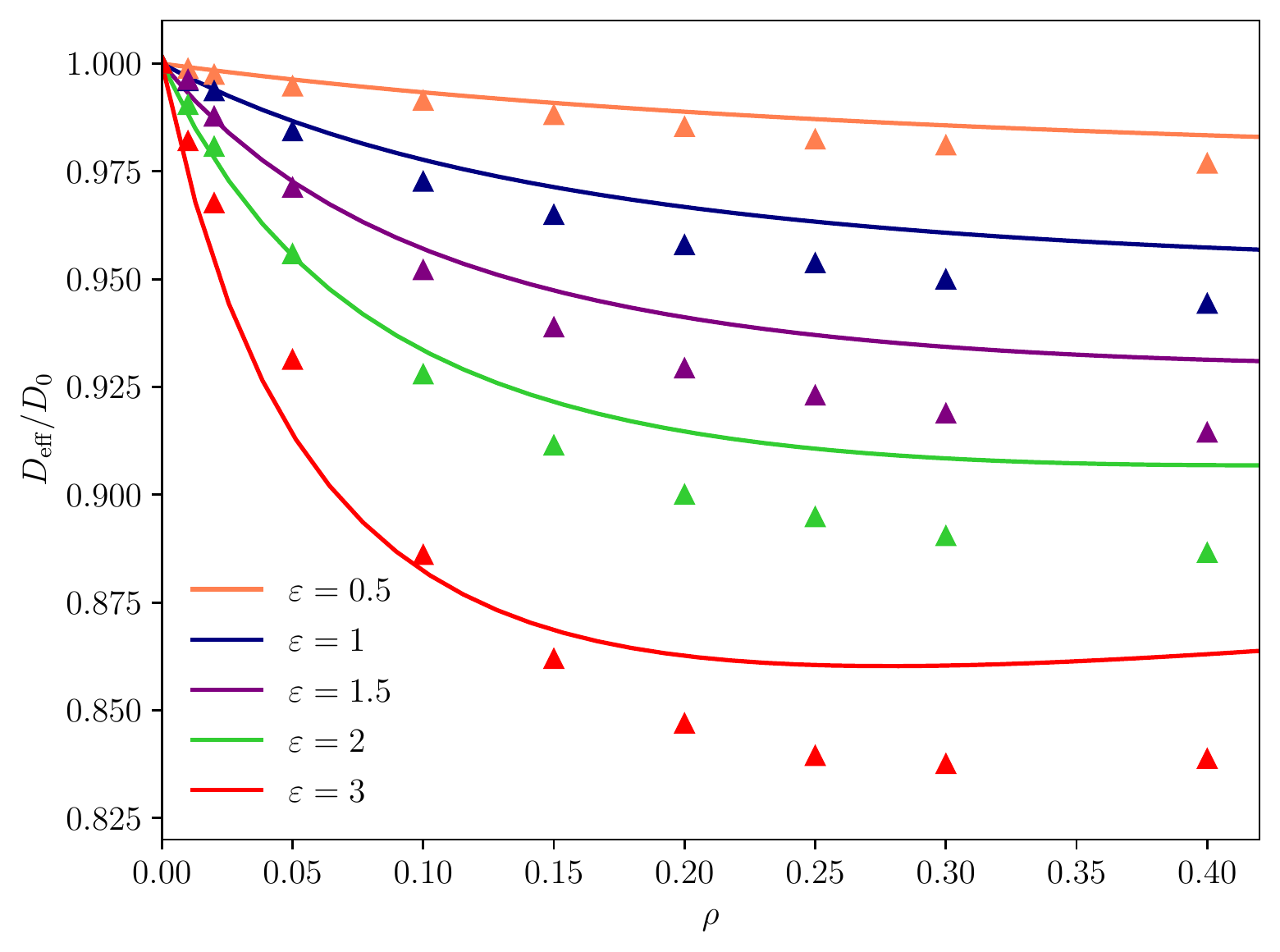}
\caption{Diffusion coefficient of a tracer as a function of the overall density $\rho$ in a suspension of identical particles interacting via a Gaussian core potential [Eq. \eqref{Gaussiancore_def}], for different values of the interaction parameter $\varepsilon$ in $\kB T$ units. The results from numerical simulations (symbols) are compared to the analytical expression obtained from our approach [Eq. \eqref{main_result_Deff}]. Error bars are within symbol size.}
\label{corps_pur}
\end{center}
\end{figure}
In order to probe the range of parameters where our approximations (linearization of the Dean-Kawasaki equation for the bath densities and limit of weak coupling between the tracer and the bath) are valid, we first consider the situation of a single component fluid, where all the particles and the tracer are of type $A$. We plot on Fig.~\ref{corps_pur} the diffusion coefficient of a tracer as a function of the density $\rho$. %For small enough densities,
In this single component fluid, the effective diffusion coefficient of a tracer is a decreasing function of the density. For a fixed value of the density, the effective diffusion coefficient decreases when the intensity of the repulsion $\varepsilon_{AA}$ increases: this is explained by the fact that crowding effects, which tend to hinder self-diffusion, are less pronounced when particles are softer. {{The relative decrease of the diffusion coefficient remains moderate in every case: for the highest value of $\varepsilon$ at the highest density, the effective diffusion coefficient is decreased by at most $17\%$.}}

The comparison between numerical simulations and the results from our analytical expression confirms its range of validity: we expect the expression of the diffusion coefficient given in Eq. \eqref{main_result_Deff} to remain valid as long as the interaction potentials are soft enough, i.e. if they remain finite and if their value at zero separation remains small or comparable to $\kB T$. Indeed, for $\varepsilon=0.5\kB T$, analytical results are in quantitative agreement with simulations even at high densities. When $\varepsilon$ increases, analytical results slightly overestimate the diffusion coefficient with a relative difference to simulation results smaller than $3$\% in the worst case.  Note that, to the best of our knowledge, the validity of the weak coupling approximation has not been investigated in this way before. Therefore,  this first comparison guides the rest of our numerical simulations, and indicates the range of parameters where our analysis is valid.

At high densities, we observe that the effective diffusion coefficient becomes an increasing function of the density. Since the particle are soft, the potential takes a finite value for $r=0$, and at high density, particle overlapping shall result in nontrivial sources of entropic increase, which may exceed the associated energetic cost. Although this effect is well-known in suspensions of soft spheres \cite{Coslovich2013,Krekelberg2009,Mausbach2006,Jacquin2010}, its relevance in the present context is not clear, and we will leave this regime aside from our analysis.

\subsection{Tracer in a binary mixture with one thermostat}

\begin{figure}
\begin{center}
\includegraphics[width=\columnwidth]{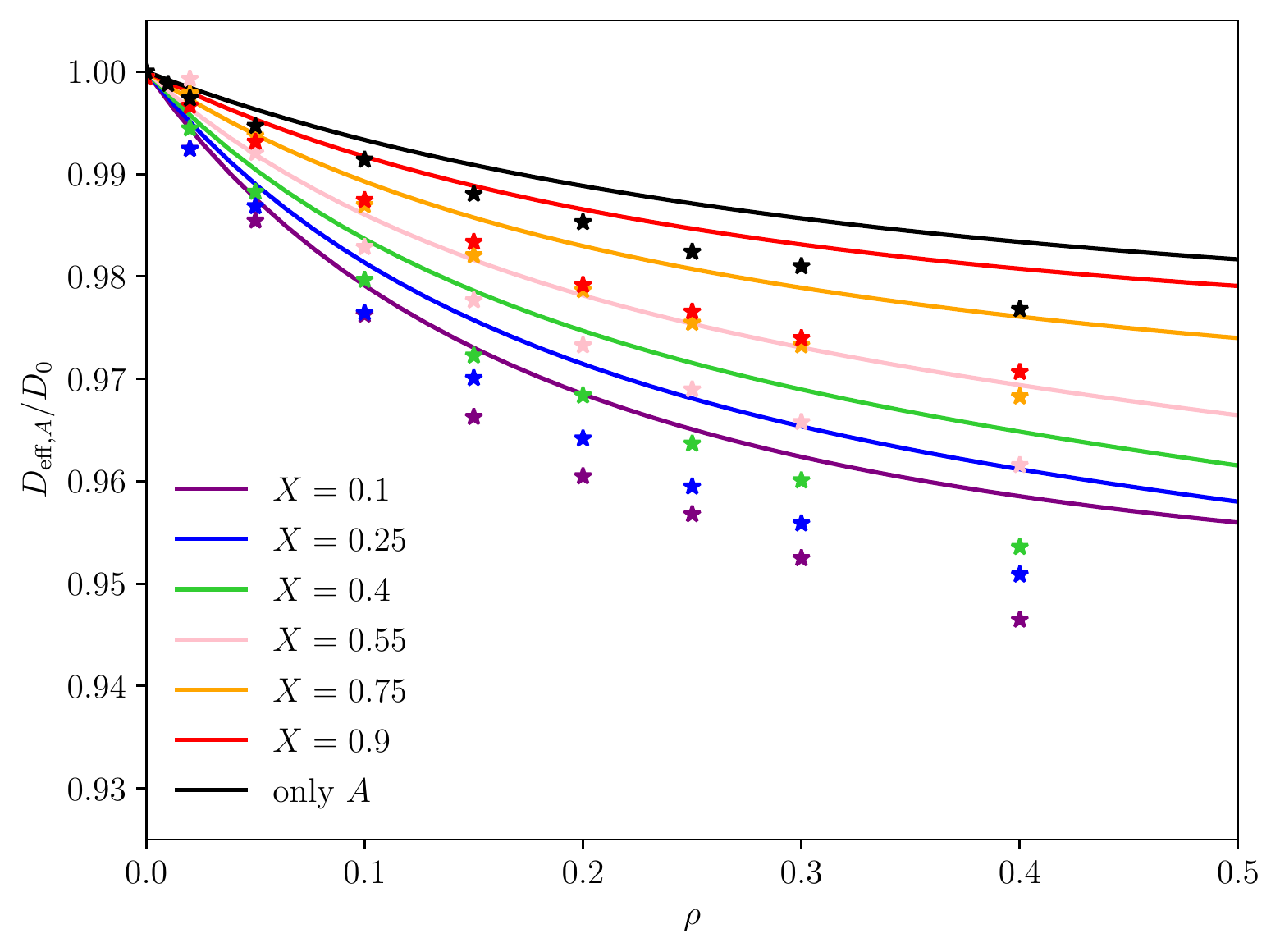}
\caption{Diffusion coefficient of particles of type $A$ as a function of the overall density $\rho$ for different compositions of the $A-B$ binary mixture ($X$ is the fraction of $A$ particles). All the particles have the same size ($\sigma_{\alpha\beta}=1$ for all $\alpha, \beta=A,B$) and are connected to the same thermostat, but  the interaction parameter $\epsilon_{\alpha\beta}$ differs depending on the considered pair: $\epsilon_{BB}=\epsilon_{AB}=\kB T$, and $\epsilon_{AA}=0.5\kB T$.}
\label{different_compositions}
\end{center}
\end{figure}

We now consider the situation of binary mixtures, made of two types of particles $A$ and $B$. We vary the fraction of $A$ particles, denoted by $X$, and we compute the effective diffusion coefficient of the $A$ particles, which play the role of tracers, divided by their bare value $D_0=D_A$. 
We assume that all the particles have the same size, but that their interaction potentials differ through the parameter $\varepsilon_{\alpha\beta}$: we choose $\varepsilon_{BB}=\varepsilon_{AB}=\kB T$, and $\varepsilon_{AA}=0.5\kB T$. In other words, the interactions between $A$ particles are softer than between the $B-B$ and $A-B$ pairs. We plot on Fig.~\ref{different_compositions} the rescaled effective diffusion coefficient of $A$ particles as a function of the overall density, both obtained from our analytical expression [Eq. \eqref{main_result_Deff}]  and from numerical simulations. We observe that, for a given value of the overall density $\rho$, the effective diffusion coefficient of soft particles decreases when the fraction of $A$ particles decreases, i.e. when the proportion of harder particles increases. This is consistent with the idea that particles tend to diffuse faster in a softer environment. The comparison between analytical and numerical results confirm that our approach provides a very good estimate of the effective diffusion coefficient of tracer in binary mixtures in the regime of weak interactions ($\varepsilon_{\alpha\beta}$ smaller or comparable to $\kB T$). Indeed, it should be noted that the difference between analytical and numerical results never exceeds $1.5\%$. { Note that the overall variation of the rescaled diffusion coefficient is of the order of 5\%}.

\subsection{Case with two different thermostats}

\begin{figure}
\begin{center}
\includegraphics[width=\columnwidth]{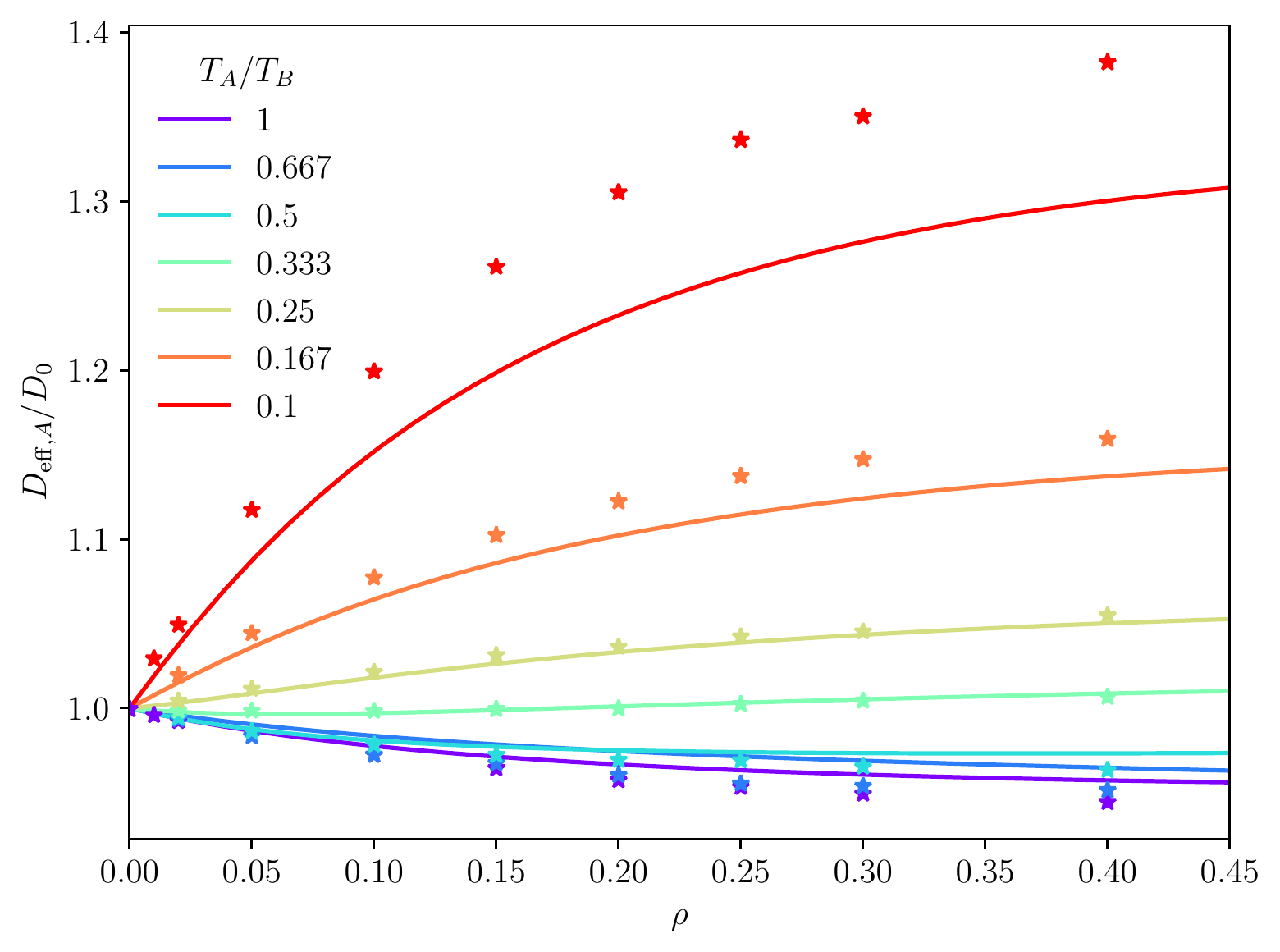}
\caption{Diffusion coefficient of particles of type $A$ in a binary $A-B$ mixture as a function of the overall density and for different values of the ratio between the temperatures of the thermostats to which each group of particle is connected. All the particles interact via the same potential ($\sigma_{\alpha\beta}=1$ and $\varepsilon_{\alpha\beta}=\kB T$ for all pairs ($\alpha,\beta$)).}
\label{different_thermostats}
\end{center}
\end{figure}

We finally consider the case of an $A-B$ binary mixture made of 5\% of $A$ particles, which play the role of tracers. All the particles interact via the same potential ($\sigma_{\alpha\beta}=1$ and $\varepsilon_{\alpha\beta}=\kB T$ for all pairs ($\alpha,\beta$)), but the two species are connected to different thermostats. We will assume that $T_A\leq T_B$, and will vary $T_A$ while maintaining $T_B$ fixed. The effective diffusion coefficient of $A$ particles as a function of the overall density for different values of the temperature ratio $T_A/T_B$ is shown on {Fig.~\ref{different_thermostats}}. When $T_A = T_B$, we retrieve the results obtained for the single component fluid (Fig.~\ref{corps_pur}). When the ratio between $T_A$ and $T_B$ decreases, i.e. when the $B$ particles become much `hotter' than the $A$ particles, the effective diffusion coefficient of the tracers with respect to their base values increases, up to a point where the enhancement induced by the `hot' bath compensates the decrease of the diffusion coefficient that results from the crowding effects (see for instance the case $T_A/T_B = 0.333$, where the rescaled effective diffusion coefficient remains very close to $1$ for all values of the density). Finally, when the two temperatures are separated by an order of magnitude (see the case $T_A/T_B = 0.1$), the crowding effects are over-compensated and the diffusion of the tracers is significantly enhanced with respect to the equilibrium reference situation: {the diffusion coefficient of the tracer is enhanced by 30\% to 40\% compared to its bare value}. Our analytical predictions are in good  agreement with numerical simulations: the difference between both is smaller than $5\%$ in every case. Finally, it should be noted that the diffusion coefficient of $B$ particles that are in large excess is almost not affected by the presence of 'colder' $A$ particles.

\section{Conclusion and perspectives}

We studied the dynamics of a tracer in contact with multiple fluctuating fields, which are not connected to the same thermostats. We derive a general analytical expression for the effective diffusion coefficient, which holds provided that the dynamics of the fluctuating media (which can represent colloidal suspensions, membranes, complex fluids...)  is linear, and that the coupling between the tracer and its environment is weak. We apply our formalism to the case of a tracer in contact with a dense binary mixture of particles which interact via soft Gaussian-core potentials, which represent polymer coils. Each type of particle is connected to a different thermostat, in such a way that one is `hot' and the other one is `cold'. Our analytical expression for the diffusion coefficient of a tracer in contact with such a mixture are confronted to Brownian dynamics simulations and are found in very good agreement. We show how the diffusivity of the tracer is affected by the heterogeneity of the mixture, and by the relative temperature of the two thermostats, therefore extending to higher densities and to different kind of interactions potentials results that were recently derived in the low-density limit and for purely repulsive hardcore interactions between particles. {The present work can be extended in multiple directions. In particular, the generality of the present formalism can be applied to study the diffusion of tracers in contact with other type of mixtures, such as electrolytes or charged media. Another natural extension would to consider the situation where the tracer is `active', for instance driven by colored noise or modeled by a run-and-tumble process.}

%\clearpage
\appendix
%\onecolumngrid

\section{Derivation of the diffusion coefficient}
\label{app_Deff}

Starting from Eqs. \eqref{sol_phi_Fourier} and \eqref{eq_tracer_lin}, the goal of this appendix is to derive the analytical expression for the effective diffusion coefficient of the tracer which is given in the main text in Eq. \eqref{main_result_Deff}.

\subsection{Generalized Langevin equation}

\emph{Operator formalism.---} Following \cite{Demery2011}, it is convenient to rewrite the evolution equation of the position of tracer $\rr_0(t)$ [Eq. \eqref{eq_tracer_lin}] and of the density fields $\phi_\alpha$ [Eq. \eqref{sol_phi_Fourier}]  under the form:
  \begin{eqnarray}
  \frac{\dd }{\dd t} \rr_0(t) &=& -\kapx \frac{\delta \mathcal{H}}{\delta \rr_0(t)}+\sqrt{\kapx} \boldsymbol{\eta}(t), \label{operator_formalism_1}\\
\partial_t \phi_\alpha(\xx,t) & =&  -\kapphi{\alpha} R_\alpha \frac{\delta \mathcal{H}}{\delta \phi_\alpha(\xx,t)}+\sqrt{\kapphi{\alpha}}\xi_\alpha (\xx,t), \label{operator_formalism_2}
\end{eqnarray}
where we introduce the following Hamiltonian, which depends on all the fields $\phi_1,\dots,\phi_\mathcal{N}$ and on the position of the tracer:
  \begin{equation}
\label{def_Hamiltonian}
\mathcal{H} = \frac{1}{2} \sum_{\alpha,\beta} \int \dd \xx \; \phi_\alpha(\xx) \Delta_{\alpha\beta} \phi_\beta(\xx) - \sum_\alpha h_\alpha K_\alpha \phi_\alpha[\rr_0(t)],
\end{equation}
and where the noise terms obey
\begin{eqnarray}
\label{ }
\moy{\eta_i(t)\eta_j(s)}& = &2\kB T_0 \delta_{ij} \delta (t-s) ,\\
\moy{\xi_\alpha(\xx,t)\xi_\beta(\xx',s)}& = &2\kB T_\alpha \delta_{\alpha\beta} R_\alpha(\xx-\xx')\delta(t-s).\nonumber\\
\end{eqnarray}

The quantities $K_\alpha$, $R_\alpha$ and $\Delta_{\alpha\beta}$ are linear operators, and we used the following shorthand notations for given operators $A$, $B$ and field $\psi$:
\begin{align}
\label{ }
A\psi (\xx) &= {\int \dd \xx' \; A(\xx-\xx') \psi(\xx'),} \\
AB\psi(\xx) &= \int \dd \xx'\int \dd \xx''  \; A(\xx-\xx')B(\xx'-\xx'') \psi(\xx''). 
\end{align}
In Fourier space, Eqs. \eqref{sol_phi_Fourier} and \eqref{eq_tracer_lin} are retrieved from the general equations \eqref{operator_formalism_1} and \eqref{operator_formalism_2} with the following relations between the operators $K_\alpha$, $R_\alpha$ and $\Delta_{\alpha\beta}$ and the pair interaction potentials $\tilde{v}_{\alpha\beta}$:
\begin{align}
&\tilde{R}_\alpha(\kk) = \tilde{R}_\beta(\kk) = k^2   \\
&\begin{cases}
 \tilde{\Delta}_{AA} = \kB T_A + X\tilde{v}_{AA} \\
 \tilde{\Delta}_{BB} = \kB T_B + (1-X)\tilde{v}_{BB}  \\
\frac{1}{2}(  \tilde{\Delta}_{AB} +\tilde{\Delta}_{BA})  = \sqrt{X (1-X)}\tilde{v}_{AB}  
\end{cases} \\
&\begin{cases}
  h_A \tilde{K}_A = \sqrt{\frac{X}{\bar{\rho}}} \tilde{v}_{A0} \\
  h_B \tilde{K}_B = \sqrt{\frac{1-X}{\bar{\rho}}} \tilde{v}_{B0} 
\end{cases}
\label{mapping}
\end{align}

\emph{Dynamics of the fields $\phi_\alpha$.---} The next step of the calculation consists in deriving a generalized Langevin equation obeyed by the position of the tracer. To this end, we first solve for the dynamics of the fields $\phi_\alpha(\xx,t)$. We start from Eq. \eqref{operator_formalism_2}, which reads in the case of a binary mixture,
 \begin{align}
\partial_t \phi_A(\xx,t)  =&  -\kapphi{A} R_A \left[ \Delta_{AA} \phi_A + \frac{1}{2} \left(  \Delta_{AB}+ \Delta_{BA}\right)\phi_{B}\right] \nonumber\\
&+h_A \kapphi{A} R_A K_A[\xx-\rr_0(t)] + \sqrt{\kapphi{A}}\xi_A(\xx,t),
\label{phi_A_real}
\end{align}
and the equivalent for $\phi_B$. The equations for $\phi_A$ and $\phi_B$ read, in Fourier space:
\begin{align}
\label{Fourier1}
&\frac{\dd}{\dd t} \begin{pmatrix}
      \widetilde{\phi}_A(\kk,t)    \\
         \widetilde{\phi}_B(\kk,t) 
\end{pmatrix}
=
-\boldsymbol{m}
\begin{pmatrix}
      \widetilde{\phi}_A   \\
         \widetilde{\phi}_B
\end{pmatrix}\nonumber \\
&+
\begin{pmatrix}
      h_A \kapphi{A}\ex{-\ii \kk \cdot \rr_0(t)}  \widetilde{R}_A  \widetilde{K}_A  + \sqrt{\kapphi{A}} \widetilde{\xi}_A \\
      h_B\kapphi{B}\ex{-\ii \kk \cdot \rr_0(t)}  \widetilde{R}_B  \widetilde{K}_B  + \sqrt{\kapphi{B}} \widetilde{\xi}_B 
\end{pmatrix},
\end{align}
where the dependences over $\kk$ are not written explicitly for clarity, and where we define the matrix $\boldsymbol{m}$ as
\begin{equation}
\boldsymbol{m} = \begin{pmatrix}
    \kapphi{A} \widetilde{R}_A  \widetilde{\Delta}_{AA} & \frac{1}{2}    \kapphi{A} \widetilde{R}_A  (\widetilde{\Delta}_{AB} +\widetilde{\Delta}_{BA} ) \\
    \frac{1}{2}    \kapphi{B} \widetilde{R}_B  (\widetilde{\Delta}_{AB} +\widetilde{\Delta}_{BA} )   &   \kapphi{B} \widetilde{R}_B  \widetilde{\Delta}_{BB}
\end{pmatrix}.
\label{expression_m}
\end{equation}
Eq. \eqref{Fourier1} is a simple set of couple linear first order differential equation, whose resolution requires the matrix exponential  $\widetilde{\boldsymbol{\mathcal{M}} } \equiv  \exp[-(t-s){\boldsymbol{m}}] $, which is written under the form
\begin{equation}
\label{matrix_exp}
\mathcal{M}_{\alpha\beta} = c^{(+)}_{\alpha\beta} \ex{-(t-s)\mu_+}+c^{(-)}_{\alpha\beta}  \ex{-(t-s)\mu_-},
\end{equation}
where we defined the matrices,
\begin{equation}
\label{c_matrices}
\boldsymbol{c^{(\pm)} } =
\frac{1}{2s}
\begin{pmatrix}
{\pm m_{AA}\mp m_{BB}+s}
 &\pm 2{m_{AB}} \\
\pm 2{m_{BA}}
   & \mp m_{AA}\pm m_{BB}+s
\end{pmatrix},
\end{equation}
the eigenvalues
\begin{equation}
\mu_\pm =  \frac{m_{AA}+m_{BB}}{2} \pm \frac{1}{2}\sqrt{(m_{AA}-m_{BB})^2+4m_{AB}m_{BA}},
\end{equation}
and the quantity
\begin{equation}
s = \sqrt{(m_{AA}-m_{BB})^2+4m_{AB}m_{BA}}.
\end{equation}
After Fourier inversion, one finds the solution of Eq. \eqref{phi_A_real} in real space under the form 
\begin{align}
\label{sol_phi_alpha_real}
\phi_\alpha(\xx,t) = &\int_{-\infty}^t \dd s \sum_\beta \left\{ h_\beta \kapphi{\beta}\mathcal{M}_{\alpha\beta}(t-s)R_\beta K_\beta [\xx-\rr_0(s)] \right. \nonumber\\
&\left. + \sqrt{\kapphi{\beta}} \mathcal{M}_{\alpha\beta}(t-s) \xi_\beta(\xx,s) \right\}
\end{align}
where $\mathcal{M}_{\alpha\beta}$ are the elements of the inverse Fourier transform of $\widetilde{\boldsymbol{\mathcal{M}} }$.\\

\emph{Dynamics of the tracer.---} Starting from Eq. \eqref{operator_formalism_1}, the dynamics of the tracer is given by
  \begin{equation}
\frac{\dd }{\dd t} \rr_0(t) = \kapx \sum_{\alpha}h_\alpha \nabla K_\alpha \phi_\alpha[\rr_0(t)] +\sqrt{\kapx} \boldsymbol{\eta}(t) 
\label{dyn_tracer}
\end{equation}
Using the expression for the field derived previously [Eq. \eqref{sol_phi_alpha_real}], the equation for the dynamics of the tracer can be rewritten as
\begin{equation}
\label{dyntracer2}
\frac{\dd}{\dd t} \rr_0(t) = \sqrt{\kapx} \boldsymbol{\eta}(t) + \int_{-\infty}^t \dd s \;  \boldsymbol{F}[\rr_0(t)-\rr_0(s),t-s]+ \boldsymbol{\Xi}[\xx,t],
\end{equation}
with
\begin{equation}
\label{ }
 \boldsymbol{F}(\xx,u)=\kapx \sum_{\alpha,\beta} h_\alpha h_\beta \kapphi{\beta} \nabla K_\alpha \mathcal{M}_{\alpha\beta} (u) R_\beta K_\beta (\xx),
\end{equation}
and
\begin{equation}
\label{ }
\boldsymbol{\Xi}[\xx,t] =\kapx \sum_{\alpha,\beta}h_\alpha \sqrt{\kapphi{\beta}} \nabla K_\alpha \int_{-\infty}^t \dd s\; \mathcal{M}_{\alpha\beta}(t-s) \xi_\beta(\xx,s)
\end{equation}

\subsection{Path-integral representation}

Starting from Eq. \eqref{dyntracer2}, we now aim at calculating the mean-square displacement of the tracer at a given time $t_f$, defined as $\langle [\rr_0(t_f)-\rr_0(0)]^2 \rangle$, and the self-diffusion coefficient, defined in Eq. \eqref{definition_Deff}. To this end, we follow the lines of Ref. \cite{Demery2011}, in which a perturbative path-integral study was outlined. Introducing a variable $\pp$ conjugated to the position of the tracer, the partition function associated to Eq. \eqref{dyntracer2} can be written under the form 
\begin{equation}
\label{ }
Z = \int \mathcal{D}\xx\,  \mathcal{D}\pp \; \ex{-S[\xx,\pp]} 
\end{equation}
where the action $S[\xx,\pp]=S_0[\xx,\pp] + S_\text{int}[\xx,\pp]$ has the following contributions:
\begin{align}
&S_0[\xx,\pp] = - \ii \int \dd t \; p_i(t) \dot x_i(t) +D_0 \int \dd t \; p_i(t) p_i(t),\\
 &S_\text{int}[\xx,\pp]  =  \ii \int \dd t \, \dd s \;  p_i(t) F_i[\xx(t)-\xx(s),t-s] \theta (t-s) \nonumber\\
 &+ \int \dd t \, \dd s \; p_i(t) G_{ij}[\xx(t)-\xx(s),t-s] p_j(t)  \theta (t-s).
\end{align}
We used the Einstein summation convention and where $\theta$ denotes the Heaviside function. The matrix elements $G_{ij}$ are defined as 
\begin{equation}
\label{ }
G_{ij}(\xx-\xx',t-t') \equiv \langle \Xi_i(\xx,t) \Xi_j(\xx',t') \rangle,
\end{equation}
and read, in Fourier space:
\begin{align}
\label{ }
\tilde G_{ij}(\kk,t) =& 2 \kappa_0^2 k_i k_j \sum_{\alpha,\beta,\gamma}h_\alpha h_\gamma \kappa_\beta  \tilde K_\alpha \tilde K_\gamma \kB T_\beta \tilde R_\beta \nonumber\\
&\times \sum_{\nu,\epsilon=\pm1}  c_{\alpha\beta}^{(\nu)}c_{\gamma\beta}^{(\epsilon)} \frac{\ex{-\mu_\nu |t|}}{\mu_\nu + \mu_\epsilon}
\end{align}
where the sums over $\alpha$, $\beta$ and $\gamma$ run over all the constituents of the mixture, and where we use the expression of the matrix exponential $\mathcal{M}_{\alpha\beta}$ given in Eq. \eqref{matrix_exp}. Expanding in the limit where the tracer-bath interactions are small (i.e. when the interaction action $S_\text{int}$ is small compared to $S_0$) and at first nontrivial order, one gets the following expression for the mean-square displacement of the tracer:
\begin{equation}
\label{ }
\langle [\rr_0(t_f)-\rr_0(0)]^2\rangle \simeq \langle [\rr_0(t_f)-\rr_0(0)]^2\rangle_0 -I_F-I_G,
\end{equation}
where the average $\langle \dots \rangle_0$ is taken over the bare action $S_0$, and where we defined 
\begin{align}
\label{}
I_F    =& \Big \langle \ii \rr_0(t_f)^2 \int \dd t\int \dd s \; \theta(t-s) p_i(t)  \nonumber\\
& \times F_{\kk,i}[\rr_0(t)-\rr_0(s),t-s] \Big \rangle_0  \\
      \underset{t_f\to\infty}{\simeq}& 4 D_0 \int \frac{\dd^d \kk}{(2\pi)^d} \;k^2 \kappa_0  \sum_{\alpha,\beta} h_\alpha h_\beta \kappa_\beta \tilde{K}_\alpha(\kk) \tilde{K}_\beta(\kk) \tilde{R}_\beta(\kk) \nonumber\\
&   \times   \sum_{\nu=\pm1}  \frac{c^{(\nu)}_{\alpha\beta}}{(D_0k^2+\mu_\nu)^2}t_f,
\end{align}
and
\begin{align}
\label{}
I_G    =& \Big \langle   \rr_0(t_f)^2 \int \dd t\int \dd s \; \theta(t-s) p_i(t) p_j(s) \nonumber\\
&G_{\kk,ij}[\rr_0(t)-\rr_0(s),t-s] \Big\rangle_0  \\
\underset{t_f\to\infty}{\simeq}& 4  \int \frac{\dd^d \kk}{(2\pi)^d} \;k^2 \kappa_0^2  \sum_{\alpha,\beta,\gamma} h_\alpha h_\gamma \kappa_\beta \tilde{K}_\alpha(\kk) \tilde{K}_\gamma(\kk) \tilde{R}_\beta(\kk) \kB T_\beta \nonumber\\
&\sum_{\nu,\epsilon=\pm 1} \frac{c_{\alpha,\beta}^{(\nu)}c_{\gamma,\beta}^{(\epsilon)}}{\mu_\nu+\mu_\epsilon} \cdot \frac{D_0 k^2-\mu_\nu}{(D_0 k^2+\mu_\nu)^2} t_f.
\end{align}
Then, using the definition of $D_\text{eff} = \lim_{t\to\infty} \langle \rr_0(t_f)^2 \rangle /(2dt_f)$ and integrating over all Fourier modes, we write the effective diffusion coefficient under the form
\begin{equation}
\label{Deff_compact}
%\boxed{
D_\text{eff} = D_0 - \sum_{\alpha,\beta} \overline{D}_{\alpha\beta}
%}
\end{equation}
with
%\begin{empheq}[box=\fbox]{align}
\begin{widetext}
\begin{align}
\label{Dalphabeta}
 \overline{D}_{\alpha\beta} = \frac{\kappa_0 \kappa_\beta}{d}  \int \frac{\dd^d \kk}{(2\pi)^d} 
k^2 [h_\alpha \tilde{K}_\alpha(\kk)] \tilde{R}_\beta(\kk)  \sum_\gamma [ h_\gamma \tilde{K}_\gamma(\kk) ] \sum_{\nu = \pm 1}  \frac{2c^{(\nu)}_{\alpha\beta}}{(D_0 k^2 + \mu_\nu)^2}   \left[  D_0 \delta_{\gamma\beta} + \kB T_\beta \kappa_0   (D_0 k^2 - \mu_\nu)  \sum_{\epsilon=\pm1} \frac{c^{(\epsilon)}_{\gamma\beta}}{\mu_\nu + \mu_\epsilon}  \right]   
%\end{empheq}
\end{align}
\end{widetext}

Finally, using the mapping between the operators $K_\alpha$, $R_\alpha$ and $\Delta_{\alpha\beta}$ and the interaction potential between the particles in the suspension [Eq. \eqref{mapping}], one gets the expression for the diffusion coefficient of the tracer given in the main text [Eq. \eqref{main_result_Deff}].

\begin{figure}
\begin{center}
\includegraphics[width=\columnwidth]{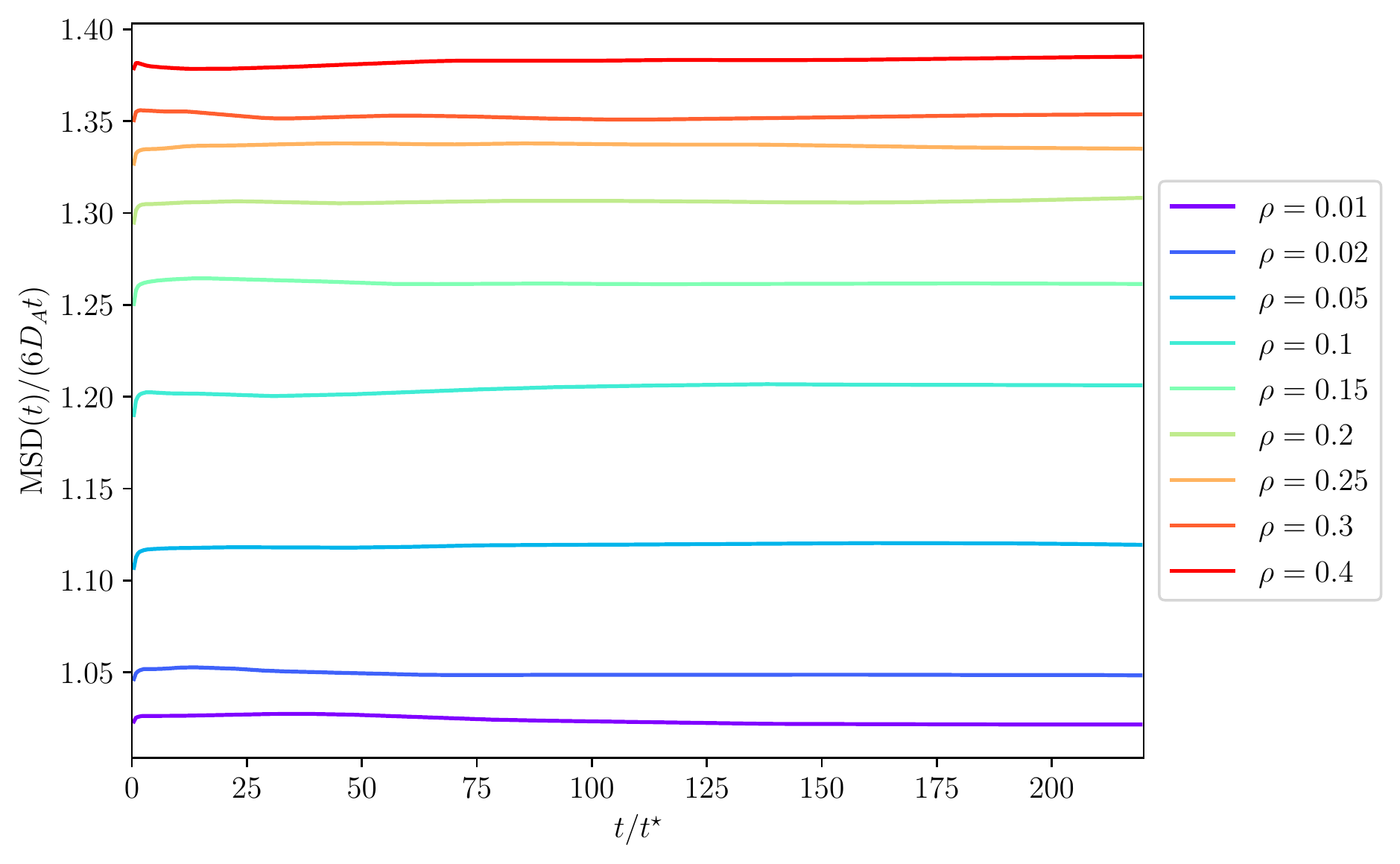}
\caption{{ Mean-squared displacements of particles of type $A$ in a binary $A-B$ mixture, divided by time, as a function of time in reduced units ($t^\star=\sigma_{BB}^2/(k_{\rm B}T_B\kappa_B)$ is the time needed for a particle $B$ to diffuse over a length equal to its size). The results obtained for several values of the overall density,  and for $T_A/T_B=0.1$ are displayed. Note that the total duration of one trajectory is here $4000 t^\star$ and that the results are averaged over $6$ independent trajectories. }}
\label{msd}
\end{center}
\end{figure}

\subsection{Convergence of the integral over Fourier modes}
\label{sec:convergence}

We finally discuss the convergence of the integral in Eq. \eqref{Dalphabeta}, along the lines of Ref. \cite{Demery2011}. Indeed, depending on the $\kk$-dependence of the operators $\tilde K_\alpha$, $\tilde R_\alpha$ and $\tilde \Delta_{\alpha \beta}$, the integral over $k=|\kk|$, may diverge, in which case it would need to be regularized by lower or upper cutoffs. The large-$k$ behavior is bounded by the typical molecular size, whereas the small-$k$ dependence is bounded by the typical system size. Depending on the small-$k$ behavior of the integrand, the integral may have a non-trivial dependence over system size, which indicates the possibility of anomalous diffusion.

Assuming that all the operators $\tilde R_\alpha$ on the one hand, and all the operators $\tilde \Delta_{\alpha \beta}$ on the other hand have identical small-$k$ behaviors, it is straightforward to show that the reasoning presented in Ref. \cite{Demery2011} still holds, and that there exists a critical dimension below which the integrals do not converge, therefore leading to anomalous diffusion.

In the present paper, we will only consider situations where diffusion is normal. In particular, in Section \ref{sec_numerical_sim}, this will be ensured by our choices of the soft interaction potentials $V_{\alpha\beta}(r)$ defined in Eq. \eqref{Gaussiancore_def}, whose Fourier transform goes to a finite constant when $k\to 0$.

\section{Details on numerical simulations}
\label{app_numerical}

To perform Brownian dynamics simulations we have used the LAMMPS computational package \cite{LAMMPS}.  We used the  command `fix Brownian' that allows one to integrate overdamped Langevin equations for the positions of particles thanks to an Euler scheme. The interaction potentials are Gaussian [Eq. \eqref{Gaussiancore_def}] and size parameters are always the same: $\sigma_{AA} = \sigma_{BB} = \sigma_{AB} = 1$. To compute interaction forces, a cutoff distance equal to $2.5\sigma_{AA}$ is used. The input mobility of particles is always the same, as it depends only on the particle size. To study the diffusion of cold $A$ tracers, we fix the temperature of the group of $A$ particles to a value smaller than that of $B$ ones. The diffusion coefficient at infinite dilution of $A$ particles is thus smaller than that of $B$ particles.

In every case, a total number of $N=4000$ particles are placed in a cubic simulation box with periodic boundary conditions. The length of the box $L_\text{box}$ is varied to change the total density $\rho$ of the system, with $\rho=N /L_\text{box}^3$. The time step is $\Delta t=0.002t^\star$, with $t^\star=\sigma_{BB}^2/(k_{\rm B}T_B\kappa_B)$ the time needed for a particle $B$ to diffuse over a length equal to its size. In each case, one long trajectory of $20\times 10^6$ time steps is first run to equilibrate the system. Then, mean squared displacements of tracers are averaged over particles and time, and also over several independent trajectories. To study the diffusion in a single component fluid, $3$ independent trajectories  of $10\times 10^6$ time steps each were done. In the case of a binary mixture with one thermostat, $3$ independent trajectories  of $20\times 10^6$ time steps each were done. To compute the tracer diffusion coefficient in mixtures with two different thermostats, results were averaged over $6$ to $9$ independent trajectories  of $20\times 10^6$ time steps each, depending on the temperature of the tracers. Cold particles are indeed intrinsically slower than hot ones, and long trajectories must be run to ensure that they travel across the whole simulation box. The uncertainty of the computed self-diffusion coefficients was evaluated from the standard deviation of values obtained from different trajectories. The uncertainty on $\frac{D_\text{eff}}{D_0}$ was in each case smaller than $0.005$. Note that the size of the symbols used in the figures is larger than these error bars. { The mean-squared displacements were found to be linear at all time for every system investigated here. An example of the results obtained for the binary $A-B$ mixture  with particles connected to two different thermostats is displayed on Fig. \ref{msd}. }

\twocolumngrid

%\bibliographystyle{apsrev4-1}
%\bibliography{la_biblio}

\begin{thebibliography}{49}%
\makeatletter
\providecommand \@ifxundefined [1]{%
 \@ifx{#1\undefined}
}%
\providecommand \@ifnum [1]{%
 \ifnum #1\expandafter \@firstoftwo
 \else \expandafter \@secondoftwo
 \fi
}%
\providecommand \@ifx [1]{%
 \ifx #1\expandafter \@firstoftwo
 \else \expandafter \@secondoftwo
 \fi
}%
\providecommand \natexlab [1]{#1}%
\providecommand \enquote  [1]{``#1''}%
\providecommand \bibnamefont  [1]{#1}%
\providecommand \bibfnamefont [1]{#1}%
\providecommand \citenamefont [1]{#1}%
\providecommand \href@noop [0]{\@secondoftwo}%
\providecommand \href [0]{\begingroup \@sanitize@url \@href}%
\providecommand \@href[1]{\@@startlink{#1}\@@href}%
\providecommand \@@href[1]{\endgroup#1\@@endlink}%
\providecommand \@sanitize@url [0]{\catcode `\\12\catcode `\$12\catcode
  `\&12\catcode `\#12\catcode `\^12\catcode `\_12\catcode `\%12\relax}%
\providecommand \@@startlink[1]{}%
\providecommand \@@endlink[0]{}%
\providecommand \url  [0]{\begingroup\@sanitize@url \@url }%
\providecommand \@url [1]{\endgroup\@href {#1}{\urlprefix }}%
\providecommand \urlprefix  [0]{URL }%
\providecommand \Eprint [0]{\href }%
\providecommand \doibase [0]{http://dx.doi.org/}%
\providecommand \selectlanguage [0]{\@gobble}%
\providecommand \bibinfo  [0]{\@secondoftwo}%
\providecommand \bibfield  [0]{\@secondoftwo}%
\providecommand \translation [1]{[#1]}%
\providecommand \BibitemOpen [0]{}%
\providecommand \bibitemStop [0]{}%
\providecommand \bibitemNoStop [0]{.\EOS\space}%
\providecommand \EOS [0]{\spacefactor3000\relax}%
\providecommand \BibitemShut  [1]{\csname bibitem#1\endcsname}%
\let\auto@bib@innerbib\@empty
%</preamble>
\bibitem [{\citenamefont {Vicsek}\ and\ \citenamefont
  {Zafeiris}(2012)}]{Vicsek2012}%
  \BibitemOpen
  \bibfield  {author} {\bibinfo {author} {\bibfnamefont {T.}~\bibnamefont
  {Vicsek}}\ and\ \bibinfo {author} {\bibfnamefont {A.}~\bibnamefont
  {Zafeiris}},\ }\href {\doibase 10.1016/j.physrep.2012.03.004} {\bibfield
  {journal} {\bibinfo  {journal} {Physics Reports}\ }\textbf {\bibinfo {volume}
  {517}},\ \bibinfo {pages} {71} (\bibinfo {year} {2012})}\BibitemShut
  {NoStop}%
\bibitem [{\citenamefont {Marchetti}\ \emph {et~al.}(2013)\citenamefont
  {Marchetti}, \citenamefont {Joanny}, \citenamefont {Ramaswamy}, \citenamefont
  {Liverpool}, \citenamefont {Prost}, \citenamefont {Rao},\ and\ \citenamefont
  {Simha}}]{Marchetti2013}%
  \BibitemOpen
  \bibfield  {author} {\bibinfo {author} {\bibfnamefont {M.~C.}\ \bibnamefont
  {Marchetti}}, \bibinfo {author} {\bibfnamefont {J.~F.}\ \bibnamefont
  {Joanny}}, \bibinfo {author} {\bibfnamefont {S.}~\bibnamefont {Ramaswamy}},
  \bibinfo {author} {\bibfnamefont {T.~B.}\ \bibnamefont {Liverpool}}, \bibinfo
  {author} {\bibfnamefont {J.}~\bibnamefont {Prost}}, \bibinfo {author}
  {\bibfnamefont {M.}~\bibnamefont {Rao}}, \ and\ \bibinfo {author}
  {\bibfnamefont {R.~A.}\ \bibnamefont {Simha}},\ }\href {\doibase
  10.1103/RevModPhys.85.1143} {\bibfield  {journal} {\bibinfo  {journal} {Rev.
  Mod. Phys.}\ }\textbf {\bibinfo {volume} {85}},\ \bibinfo {pages} {1143}
  (\bibinfo {year} {2013})}\BibitemShut {NoStop}%
\bibitem [{\citenamefont {Cates}\ and\ \citenamefont
  {Tailleur}(2015)}]{Cates2015}%
  \BibitemOpen
  \bibfield  {author} {\bibinfo {author} {\bibfnamefont {M.~E.}\ \bibnamefont
  {Cates}}\ and\ \bibinfo {author} {\bibfnamefont {J.}~\bibnamefont
  {Tailleur}},\ }\href {\doibase 10.1146/annurev-conmatphys-031214-014710}
  {\bibfield  {journal} {\bibinfo  {journal} {Ann. Rev. Condens. Matter Phys.}\
  }\textbf {\bibinfo {volume} {6}},\ \bibinfo {pages} {219} (\bibinfo {year}
  {2015})}\BibitemShut {NoStop}%
\bibitem [{\citenamefont {Bechinger}\ \emph {et~al.}(2016)\citenamefont
  {Bechinger}, \citenamefont {{Di Leonardo}}, \citenamefont {L{\"{o}}wen},
  \citenamefont {Reichhardt}, \citenamefont {Volpe},\ and\ \citenamefont
  {Volpe}}]{Bechinger2016}%
  \BibitemOpen
  \bibfield  {author} {\bibinfo {author} {\bibfnamefont {C.}~\bibnamefont
  {Bechinger}}, \bibinfo {author} {\bibfnamefont {R.}~\bibnamefont {{Di
  Leonardo}}}, \bibinfo {author} {\bibfnamefont {H.}~\bibnamefont
  {L{\"{o}}wen}}, \bibinfo {author} {\bibfnamefont {C.}~\bibnamefont
  {Reichhardt}}, \bibinfo {author} {\bibfnamefont {G.}~\bibnamefont {Volpe}}, \
  and\ \bibinfo {author} {\bibfnamefont {G.}~\bibnamefont {Volpe}},\ }\href
  {\doibase 10.1103/RevModPhys.88.045006} {\bibfield  {journal} {\bibinfo
  {journal} {Rev. Mod. Phys.}\ }\textbf {\bibinfo {volume} {88}},\ \bibinfo
  {pages} {045006} (\bibinfo {year} {2016})}\BibitemShut {NoStop}%
\bibitem [{\citenamefont {Guo}\ \emph {et~al.}(2014)\citenamefont {Guo},
  \citenamefont {Ehrlicher}, \citenamefont {Jensen}, \citenamefont {Renz},
  \citenamefont {Moore}, \citenamefont {Goldman}, \citenamefont
  {Lippincott-Schwartz}, \citenamefont {Mackintosh},\ and\ \citenamefont
  {Weitz}}]{Guo2014}%
  \BibitemOpen
  \bibfield  {author} {\bibinfo {author} {\bibfnamefont {M.}~\bibnamefont
  {Guo}}, \bibinfo {author} {\bibfnamefont {A.~J.}\ \bibnamefont {Ehrlicher}},
  \bibinfo {author} {\bibfnamefont {M.~H.}\ \bibnamefont {Jensen}}, \bibinfo
  {author} {\bibfnamefont {M.}~\bibnamefont {Renz}}, \bibinfo {author}
  {\bibfnamefont {J.~R.}\ \bibnamefont {Moore}}, \bibinfo {author}
  {\bibfnamefont {R.~D.}\ \bibnamefont {Goldman}}, \bibinfo {author}
  {\bibfnamefont {J.}~\bibnamefont {Lippincott-Schwartz}}, \bibinfo {author}
  {\bibfnamefont {F.~C.}\ \bibnamefont {Mackintosh}}, \ and\ \bibinfo {author}
  {\bibfnamefont {D.~A.}\ \bibnamefont {Weitz}},\ }\href {\doibase
  10.1016/j.cell.2014.06.051} {\bibfield  {journal} {\bibinfo  {journal}
  {Cell}\ }\textbf {\bibinfo {volume} {158}},\ \bibinfo {pages} {822} (\bibinfo
  {year} {2014})}\BibitemShut {NoStop}%
\bibitem [{\citenamefont {Parry}\ \emph {et~al.}(2014)\citenamefont {Parry},
  \citenamefont {Surovtsev}, \citenamefont {Cabeen}, \citenamefont {O'Hern},
  \citenamefont {Dufresne},\ and\ \citenamefont {Jacobs-Wagner}}]{Parry2014}%
  \BibitemOpen
  \bibfield  {author} {\bibinfo {author} {\bibfnamefont {B.~R.}\ \bibnamefont
  {Parry}}, \bibinfo {author} {\bibfnamefont {I.~V.}\ \bibnamefont
  {Surovtsev}}, \bibinfo {author} {\bibfnamefont {M.~T.}\ \bibnamefont
  {Cabeen}}, \bibinfo {author} {\bibfnamefont {C.~S.}\ \bibnamefont {O'Hern}},
  \bibinfo {author} {\bibfnamefont {E.~R.}\ \bibnamefont {Dufresne}}, \ and\
  \bibinfo {author} {\bibfnamefont {C.}~\bibnamefont {Jacobs-Wagner}},\ }\href
  {\doibase 10.1016/j.cell.2013.11.028} {\bibfield  {journal} {\bibinfo
  {journal} {Cell}\ }\textbf {\bibinfo {volume} {156}},\ \bibinfo {pages} {183}
  (\bibinfo {year} {2014})}\BibitemShut {NoStop}%
\bibitem [{\citenamefont {Weber}\ \emph {et~al.}(2016)\citenamefont {Weber},
  \citenamefont {Weber},\ and\ \citenamefont {Frey}}]{Weber2016}%
  \BibitemOpen
  \bibfield  {author} {\bibinfo {author} {\bibfnamefont {S.~N.}\ \bibnamefont
  {Weber}}, \bibinfo {author} {\bibfnamefont {C.~A.}\ \bibnamefont {Weber}}, \
  and\ \bibinfo {author} {\bibfnamefont {E.}~\bibnamefont {Frey}},\ }\href
  {\doibase 10.1103/PhysRevLett.116.058301} {\bibfield  {journal} {\bibinfo
  {journal} {Phys. Rev. Lett.}\ }\textbf {\bibinfo {volume} {116}},\ \bibinfo
  {pages} {058301} (\bibinfo {year} {2016})}\BibitemShut {NoStop}%
\bibitem [{\citenamefont {Tanaka}\ \emph {et~al.}(2017)\citenamefont {Tanaka},
  \citenamefont {Lee},\ and\ \citenamefont {Brenner}}]{Tanaka2016a}%
  \BibitemOpen
  \bibfield  {author} {\bibinfo {author} {\bibfnamefont {H.}~\bibnamefont
  {Tanaka}}, \bibinfo {author} {\bibfnamefont {A.~A.}\ \bibnamefont {Lee}}, \
  and\ \bibinfo {author} {\bibfnamefont {M.~P.}\ \bibnamefont {Brenner}},\
  }\href {\doibase 10.1103/PhysRevFluids.2.043103} {\bibfield  {journal}
  {\bibinfo  {journal} {Phys. Rev. Fluids}\ }\textbf {\bibinfo {volume} {2}},\
  \bibinfo {pages} {043103} (\bibinfo {year} {2017})}\BibitemShut {NoStop}%
\bibitem [{\citenamefont {Smrek}\ and\ \citenamefont
  {Kremer}(2018)}]{Smrek2018}%
  \BibitemOpen
  \bibfield  {author} {\bibinfo {author} {\bibfnamefont {J.}~\bibnamefont
  {Smrek}}\ and\ \bibinfo {author} {\bibfnamefont {K.}~\bibnamefont {Kremer}},\
  }\href {\doibase 10.3390/e20070520} {\bibfield  {journal} {\bibinfo
  {journal} {Entropy}\ }\textbf {\bibinfo {volume} {20}},\ \bibinfo {pages}
  {520} (\bibinfo {year} {2018})}\BibitemShut {NoStop}%
\bibitem [{\citenamefont {Smrek}\ and\ \citenamefont
  {Kremer}(2017)}]{Smrek2017}%
  \BibitemOpen
  \bibfield  {author} {\bibinfo {author} {\bibfnamefont {J.}~\bibnamefont
  {Smrek}}\ and\ \bibinfo {author} {\bibfnamefont {K.}~\bibnamefont {Kremer}},\
  }\href {\doibase 10.1103/PhysRevLett.118.098002} {\bibfield  {journal}
  {\bibinfo  {journal} {Physical Review Letters}\ }\textbf {\bibinfo {volume}
  {118}},\ \bibinfo {pages} {098002} (\bibinfo {year} {2017})}\BibitemShut
  {NoStop}%
\bibitem [{\citenamefont {Chubak}\ \emph {et~al.}(2020)\citenamefont {Chubak},
  \citenamefont {Likos}, \citenamefont {Kremer},\ and\ \citenamefont
  {Smrek}}]{Chubak2020}%
  \BibitemOpen
  \bibfield  {author} {\bibinfo {author} {\bibfnamefont {I.}~\bibnamefont
  {Chubak}}, \bibinfo {author} {\bibfnamefont {C.~N.}\ \bibnamefont {Likos}},
  \bibinfo {author} {\bibfnamefont {K.}~\bibnamefont {Kremer}}, \ and\ \bibinfo
  {author} {\bibfnamefont {J.}~\bibnamefont {Smrek}},\ }\href {\doibase
  10.1103/PhysRevResearch.2.043249} {\bibfield  {journal} {\bibinfo  {journal}
  {Physical Review Research}\ }\textbf {\bibinfo {volume} {2}},\ \bibinfo
  {pages} {43249} (\bibinfo {year} {2020})}\BibitemShut {NoStop}%
\bibitem [{\citenamefont {Smrek}\ \emph {et~al.}(2020)\citenamefont {Smrek},
  \citenamefont {Chubak}, \citenamefont {Likos},\ and\ \citenamefont
  {Kremer}}]{Smrek2020}%
  \BibitemOpen
  \bibfield  {author} {\bibinfo {author} {\bibfnamefont {J.}~\bibnamefont
  {Smrek}}, \bibinfo {author} {\bibfnamefont {I.}~\bibnamefont {Chubak}},
  \bibinfo {author} {\bibfnamefont {C.~N.}\ \bibnamefont {Likos}}, \ and\
  \bibinfo {author} {\bibfnamefont {K.}~\bibnamefont {Kremer}},\ }\href
  {\doibase 10.1038/s41467-019-13696-z} {\bibfield  {journal} {\bibinfo
  {journal} {Nature Communications}\ }\textbf {\bibinfo {volume} {11}},\
  \bibinfo {pages} {26} (\bibinfo {year} {2020})}\BibitemShut {NoStop}%
\bibitem [{\citenamefont {Lu}\ \emph {et~al.}(2015)\citenamefont {Lu},
  \citenamefont {Dean},\ and\ \citenamefont {Podgornik}}]{Dean}%
  \BibitemOpen
  \bibfield  {author} {\bibinfo {author} {\bibfnamefont {B.~S.}\ \bibnamefont
  {Lu}}, \bibinfo {author} {\bibfnamefont {D.~S.}\ \bibnamefont {Dean}}, \ and\
  \bibinfo {author} {\bibfnamefont {R.}~\bibnamefont {Podgornik}},\ }\href
  {\doibase 10.1209/0295-5075/112/20001} {\bibfield  {journal} {\bibinfo
  {journal} {Europhys. Lett.}\ }\textbf {\bibinfo {volume} {112}},\ \bibinfo
  {pages} {20001} (\bibinfo {year} {2015})}\BibitemShut {NoStop}%
\bibitem [{\citenamefont {Grosberg}\ and\ \citenamefont
  {Joanny}(2015)}]{Grosberg2015}%
  \BibitemOpen
  \bibfield  {author} {\bibinfo {author} {\bibfnamefont {A.~Y.}\ \bibnamefont
  {Grosberg}}\ and\ \bibinfo {author} {\bibfnamefont {J.~F.}\ \bibnamefont
  {Joanny}},\ }\href {\doibase 10.1103/PhysRevE.92.032118} {\bibfield
  {journal} {\bibinfo  {journal} {Phys. Rev. E}\ }\textbf {\bibinfo {volume}
  {92}},\ \bibinfo {pages} {032118} (\bibinfo {year} {2015})}\BibitemShut
  {NoStop}%
\bibitem [{\citenamefont {Ilker}\ and\ \citenamefont
  {Joanny}(2020)}]{Ilker2020}%
  \BibitemOpen
  \bibfield  {author} {\bibinfo {author} {\bibfnamefont {E.}~\bibnamefont
  {Ilker}}\ and\ \bibinfo {author} {\bibfnamefont {J.-F.}\ \bibnamefont
  {Joanny}},\ }\href {\doibase 10.1103/physrevresearch.2.023200} {\bibfield
  {journal} {\bibinfo  {journal} {Physical Review Research}\ }\textbf {\bibinfo
  {volume} {2}},\ \bibinfo {pages} {23200} (\bibinfo {year}
  {2020})}\BibitemShut {NoStop}%
\bibitem [{\citenamefont {Wang}\ and\ \citenamefont
  {Grosberg}(2020)}]{Wang2020}%
  \BibitemOpen
  \bibfield  {author} {\bibinfo {author} {\bibfnamefont {M.}~\bibnamefont
  {Wang}}\ and\ \bibinfo {author} {\bibfnamefont {A.~Y.}\ \bibnamefont
  {Grosberg}},\ }\href {\doibase 10.1103/PhysRevE.101.032131} {\bibfield
  {journal} {\bibinfo  {journal} {Physical Review E}\ }\textbf {\bibinfo
  {volume} {101}},\ \bibinfo {pages} {032131} (\bibinfo {year}
  {2020})}\BibitemShut {NoStop}%
\bibitem [{\citenamefont {Ilker}\ \emph {et~al.}(2021)\citenamefont {Ilker},
  \citenamefont {Castellana},\ and\ \citenamefont {Joanny}}]{Ilker2021}%
  \BibitemOpen
  \bibfield  {author} {\bibinfo {author} {\bibfnamefont {E.}~\bibnamefont
  {Ilker}}, \bibinfo {author} {\bibfnamefont {M.}~\bibnamefont {Castellana}}, \
  and\ \bibinfo {author} {\bibfnamefont {J.-F.}\ \bibnamefont {Joanny}},\
  }\href {http://arxiv.org/abs/2103.06659} {\bibfield  {journal} {\bibinfo
  {journal} {Phys. Rev. Research}\ }\textbf {\bibinfo {volume} {3}},\ \bibinfo
  {pages} {023207} (\bibinfo {year} {2021})}\BibitemShut {NoStop}%
\bibitem [{\citenamefont {D{\'{e}}mery}\ and\ \citenamefont
  {Dean}(2011)}]{Demery2011}%
  \BibitemOpen
  \bibfield  {author} {\bibinfo {author} {\bibfnamefont {V.}~\bibnamefont
  {D{\'{e}}mery}}\ and\ \bibinfo {author} {\bibfnamefont {D.~S.}\ \bibnamefont
  {Dean}},\ }\href {\doibase 10.1103/PhysRevE.84.011148} {\bibfield  {journal}
  {\bibinfo  {journal} {Phys. Rev. E}\ }\textbf {\bibinfo {volume} {84}},\
  \bibinfo {pages} {011148} (\bibinfo {year} {2011})}\BibitemShut {NoStop}%
\bibitem [{\citenamefont {Louis}\ \emph
  {et~al.}(2000{\natexlab{a}})\citenamefont {Louis}, \citenamefont {Bolhuis},\
  and\ \citenamefont {Hansen}}]{Louis2000}%
  \BibitemOpen
  \bibfield  {author} {\bibinfo {author} {\bibfnamefont {A.~A.}\ \bibnamefont
  {Louis}}, \bibinfo {author} {\bibfnamefont {P.~G.}\ \bibnamefont {Bolhuis}},
  \ and\ \bibinfo {author} {\bibfnamefont {J.~P.}\ \bibnamefont {Hansen}},\
  }\href {\doibase 10.1103/PhysRevE.62.7961} {\bibfield  {journal} {\bibinfo
  {journal} {Phys. Rev. E}\ }\textbf {\bibinfo {volume} {62}},\ \bibinfo
  {pages} {7961} (\bibinfo {year} {2000}{\natexlab{a}})}\BibitemShut {NoStop}%
\bibitem [{\citenamefont {Lang}\ \emph {et~al.}(2000)\citenamefont {Lang},
  \citenamefont {Likos}, \citenamefont {Watzlawek},\ and\ \citenamefont
  {L{\"{o}}wen}}]{Lang2000}%
  \BibitemOpen
  \bibfield  {author} {\bibinfo {author} {\bibfnamefont {A.}~\bibnamefont
  {Lang}}, \bibinfo {author} {\bibfnamefont {C.~N.}\ \bibnamefont {Likos}},
  \bibinfo {author} {\bibfnamefont {M.}~\bibnamefont {Watzlawek}}, \ and\
  \bibinfo {author} {\bibfnamefont {H.}~\bibnamefont {L{\"{o}}wen}},\ }\href
  {\doibase 10.1088/0953-8984/12/24/302} {\bibfield  {journal} {\bibinfo
  {journal} {Journal of Physics Condensed Matter}\ }\textbf {\bibinfo {volume}
  {12}},\ \bibinfo {pages} {5087} (\bibinfo {year} {2000})}\BibitemShut
  {NoStop}%
\bibitem [{\citenamefont {Likos}\ \emph {et~al.}(2001)\citenamefont {Likos},
  \citenamefont {Lang}, \citenamefont {Watzlawek},\ and\ \citenamefont
  {L{\"{o}}wen}}]{Likos2001}%
  \BibitemOpen
  \bibfield  {author} {\bibinfo {author} {\bibfnamefont {C.~N.}\ \bibnamefont
  {Likos}}, \bibinfo {author} {\bibfnamefont {A.}~\bibnamefont {Lang}},
  \bibinfo {author} {\bibfnamefont {M.}~\bibnamefont {Watzlawek}}, \ and\
  \bibinfo {author} {\bibfnamefont {H.}~\bibnamefont {L{\"{o}}wen}},\ }\href
  {\doibase 10.1103/PhysRevE.63.031206} {\bibfield  {journal} {\bibinfo
  {journal} {Phys. Rev. E}\ }\textbf {\bibinfo {volume} {63}},\ \bibinfo
  {pages} {031206} (\bibinfo {year} {2001})}\BibitemShut {NoStop}%
\bibitem [{\citenamefont {Wensink}\ \emph {et~al.}(2008)\citenamefont
  {Wensink}, \citenamefont {L{\"{o}}wen}, \citenamefont {Rex}, \citenamefont
  {Likos},\ and\ \citenamefont {van Teeffelen}}]{Wensink2008}%
  \BibitemOpen
  \bibfield  {author} {\bibinfo {author} {\bibfnamefont {H.~H.}\ \bibnamefont
  {Wensink}}, \bibinfo {author} {\bibfnamefont {H.}~\bibnamefont
  {L{\"{o}}wen}}, \bibinfo {author} {\bibfnamefont {M.}~\bibnamefont {Rex}},
  \bibinfo {author} {\bibfnamefont {C.~N.}\ \bibnamefont {Likos}}, \ and\
  \bibinfo {author} {\bibfnamefont {S.}~\bibnamefont {van Teeffelen}},\ }\href
  {\doibase 10.1016/j.cpc.2008.01.009} {\bibfield  {journal} {\bibinfo
  {journal} {Computer Physics Communications}\ }\textbf {\bibinfo {volume}
  {179}},\ \bibinfo {pages} {77} (\bibinfo {year} {2008})}\BibitemShut
  {NoStop}%
\bibitem [{\citenamefont {Kawasaki}(1994)}]{Kawasaki1994}%
  \BibitemOpen
  \bibfield  {author} {\bibinfo {author} {\bibfnamefont {K.}~\bibnamefont
  {Kawasaki}},\ }\href {\doibase 10.1016/0378-4371(94)90533-9} {\bibfield
  {journal} {\bibinfo  {journal} {Physica A: Statistical Mechanics and its
  Applications}\ }\textbf {\bibinfo {volume} {208}},\ \bibinfo {pages} {35}
  (\bibinfo {year} {1994})}\BibitemShut {NoStop}%
\bibitem [{\citenamefont {Dean}(1996)}]{Dean1996}%
  \BibitemOpen
  \bibfield  {author} {\bibinfo {author} {\bibfnamefont {D.~S.}\ \bibnamefont
  {Dean}},\ }\href {\doibase 10.1088/0305-4470/29/24/001} {\bibfield  {journal}
  {\bibinfo  {journal} {J. Phys. A: Math. Gen.}\ }\textbf {\bibinfo {volume}
  {29}},\ \bibinfo {pages} {L613} (\bibinfo {year} {1996})}\BibitemShut
  {NoStop}%
\bibitem [{\citenamefont {D{\'{e}}mery}\ \emph {et~al.}(2014)\citenamefont
  {D{\'{e}}mery}, \citenamefont {B{\'{e}}nichou},\ and\ \citenamefont
  {Jacquin}}]{Demery2014}%
  \BibitemOpen
  \bibfield  {author} {\bibinfo {author} {\bibfnamefont {V.}~\bibnamefont
  {D{\'{e}}mery}}, \bibinfo {author} {\bibfnamefont {O.}~\bibnamefont
  {B{\'{e}}nichou}}, \ and\ \bibinfo {author} {\bibfnamefont {H.}~\bibnamefont
  {Jacquin}},\ }\href {\doibase 10.1088/1367-2630/16/5/053032} {\bibfield
  {journal} {\bibinfo  {journal} {New J. Phys.}\ }\textbf {\bibinfo {volume}
  {16}},\ \bibinfo {pages} {053032} (\bibinfo {year} {2014})}\BibitemShut
  {NoStop}%
\bibitem [{\citenamefont {D{\'{e}}mery}\ and\ \citenamefont
  {Fodor}(2019)}]{Demery2019}%
  \BibitemOpen
  \bibfield  {author} {\bibinfo {author} {\bibfnamefont {V.}~\bibnamefont
  {D{\'{e}}mery}}\ and\ \bibinfo {author} {\bibfnamefont
  {{\'{E}}.}~\bibnamefont {Fodor}},\ }\href {\doibase 10.1088/1742-5468/ab02e9}
  {\bibfield  {journal} {\bibinfo  {journal} {J. Stat. Mech}\ }\textbf
  {\bibinfo {volume} {2019}},\ \bibinfo {pages} {033202} (\bibinfo {year}
  {2019})}\BibitemShut {NoStop}%
\bibitem [{\citenamefont {D{\'{e}}mery}(2015)}]{Demery2015}%
  \BibitemOpen
  \bibfield  {author} {\bibinfo {author} {\bibfnamefont {V.}~\bibnamefont
  {D{\'{e}}mery}},\ }\href {\doibase 10.1103/PhysRevE.91.062301} {\bibfield
  {journal} {\bibinfo  {journal} {Phys. Rev. E}\ }\textbf {\bibinfo {volume}
  {91}},\ \bibinfo {pages} {062301} (\bibinfo {year} {2015})}\BibitemShut
  {NoStop}%
\bibitem [{\citenamefont {Feng}\ and\ \citenamefont {Hou}(2021)}]{Feng2021}%
  \BibitemOpen
  \bibfield  {author} {\bibinfo {author} {\bibfnamefont {M.}~\bibnamefont
  {Feng}}\ and\ \bibinfo {author} {\bibfnamefont {Z.}~\bibnamefont {Hou}},\
  }\href {http://arxiv.org/abs/2110.00279} {\bibfield  {journal} {\bibinfo
  {journal} {arXiv:2110.00279}\ } (\bibinfo {year} {2021})}\BibitemShut
  {NoStop}%
\bibitem [{\citenamefont {Poncet}\ \emph {et~al.}(2021)\citenamefont {Poncet},
  \citenamefont {B{\'{e}}nichou}, \citenamefont {D{\'{e}}mery},\ and\
  \citenamefont {Nishiguchi}}]{Poncet2021b}%
  \BibitemOpen
  \bibfield  {author} {\bibinfo {author} {\bibfnamefont {A.}~\bibnamefont
  {Poncet}}, \bibinfo {author} {\bibfnamefont {O.}~\bibnamefont
  {B{\'{e}}nichou}}, \bibinfo {author} {\bibfnamefont {V.}~\bibnamefont
  {D{\'{e}}mery}}, \ and\ \bibinfo {author} {\bibfnamefont {D.}~\bibnamefont
  {Nishiguchi}},\ }\href {\doibase 10.1103/PhysRevE.103.012605} {\bibfield
  {journal} {\bibinfo  {journal} {Physical Review E}\ }\textbf {\bibinfo
  {volume} {103}},\ \bibinfo {pages} {012605} (\bibinfo {year}
  {2021})}\BibitemShut {NoStop}%
\bibitem [{\citenamefont {Tociu}\ \emph {et~al.}(2019)\citenamefont {Tociu},
  \citenamefont {Fodor}, \citenamefont {Nemoto},\ and\ \citenamefont
  {Vaikuntanathan}}]{Tociu2019}%
  \BibitemOpen
  \bibfield  {author} {\bibinfo {author} {\bibfnamefont {L.}~\bibnamefont
  {Tociu}}, \bibinfo {author} {\bibfnamefont {{\'{E}}.}~\bibnamefont {Fodor}},
  \bibinfo {author} {\bibfnamefont {T.}~\bibnamefont {Nemoto}}, \ and\ \bibinfo
  {author} {\bibfnamefont {S.}~\bibnamefont {Vaikuntanathan}},\ }\href
  {\doibase 10.1103/PhysRevX.9.041026} {\bibfield  {journal} {\bibinfo
  {journal} {Physical Review X}\ }\textbf {\bibinfo {volume} {9}},\ \bibinfo
  {pages} {41026} (\bibinfo {year} {2019})}\BibitemShut {NoStop}%
\bibitem [{\citenamefont {Fodor}\ \emph {et~al.}(2020)\citenamefont {Fodor},
  \citenamefont {Nemoto},\ and\ \citenamefont {Vaikuntanathan}}]{Fodor2020}%
  \BibitemOpen
  \bibfield  {author} {\bibinfo {author} {\bibfnamefont {{\'{E}}.}~\bibnamefont
  {Fodor}}, \bibinfo {author} {\bibfnamefont {T.}~\bibnamefont {Nemoto}}, \
  and\ \bibinfo {author} {\bibfnamefont {S.}~\bibnamefont {Vaikuntanathan}},\
  }\href {\doibase 10.1088/1367-2630/ab6353} {\bibfield  {journal} {\bibinfo
  {journal} {New Journal of Physics}\ }\textbf {\bibinfo {volume} {22}},\
  \bibinfo {pages} {013052} (\bibinfo {year} {2020})}\BibitemShut {NoStop}%
\bibitem [{\citenamefont {Rassolov}\ \emph {et~al.}(2022)\citenamefont
  {Rassolov}, \citenamefont {Tociu}, \citenamefont {Fodor},\ and\ \citenamefont
  {Vaikuntanathan}}]{Tociu2020}%
  \BibitemOpen
  \bibfield  {author} {\bibinfo {author} {\bibfnamefont {G.}~\bibnamefont
  {Rassolov}}, \bibinfo {author} {\bibfnamefont {L.}~\bibnamefont {Tociu}},
  \bibinfo {author} {\bibfnamefont {E.}~\bibnamefont {Fodor}}, \ and\ \bibinfo
  {author} {\bibfnamefont {S.}~\bibnamefont {Vaikuntanathan}},\ }\href
  {\doibase 10.1063/5.0097863} {\bibfield  {journal} {\bibinfo  {journal} {J.
  Chem. Phys.}\ }\textbf {\bibinfo {volume} {157}},\ \bibinfo {pages} {054901}
  (\bibinfo {year} {2022})}\BibitemShut {NoStop}%
\bibitem [{\citenamefont {Martin}\ \emph {et~al.}(2018)\citenamefont {Martin},
  \citenamefont {Nardini}, \citenamefont {Cates},\ and\ \citenamefont
  {Fodor}}]{Martin2018}%
  \BibitemOpen
  \bibfield  {author} {\bibinfo {author} {\bibfnamefont {D.}~\bibnamefont
  {Martin}}, \bibinfo {author} {\bibfnamefont {C.}~\bibnamefont {Nardini}},
  \bibinfo {author} {\bibfnamefont {M.~E.}\ \bibnamefont {Cates}}, \ and\
  \bibinfo {author} {\bibnamefont {Fodor}},\ }\href {\doibase
  10.1209/0295-5075/121/60005} {\bibfield  {journal} {\bibinfo  {journal}
  {Europhys. Lett.}\ }\textbf {\bibinfo {volume} {121}},\ \bibinfo {pages}
  {60005} (\bibinfo {year} {2018})}\BibitemShut {NoStop}%
\bibitem [{\citenamefont {Poncet}\ \emph {et~al.}(2017)\citenamefont {Poncet},
  \citenamefont {B{\'{e}}nichou}, \citenamefont {D{\'{e}}mery},\ and\
  \citenamefont {Oshanin}}]{Poncet2016}%
  \BibitemOpen
  \bibfield  {author} {\bibinfo {author} {\bibfnamefont {A.}~\bibnamefont
  {Poncet}}, \bibinfo {author} {\bibfnamefont {O.}~\bibnamefont
  {B{\'{e}}nichou}}, \bibinfo {author} {\bibfnamefont {V.}~\bibnamefont
  {D{\'{e}}mery}}, \ and\ \bibinfo {author} {\bibfnamefont {G.}~\bibnamefont
  {Oshanin}},\ }\href {\doibase 10.1103/PhysRevLett.118.118002} {\bibfield
  {journal} {\bibinfo  {journal} {Phys. Rev. Lett.}\ }\textbf {\bibinfo
  {volume} {118}},\ \bibinfo {pages} {118002} (\bibinfo {year}
  {2017})}\BibitemShut {NoStop}%
\bibitem [{\citenamefont {Mahdisoltani}\ and\ \citenamefont
  {Golestanian}(2021{\natexlab{a}})}]{Mahdisoltani2021a}%
  \BibitemOpen
  \bibfield  {author} {\bibinfo {author} {\bibfnamefont {S.}~\bibnamefont
  {Mahdisoltani}}\ and\ \bibinfo {author} {\bibfnamefont {R.}~\bibnamefont
  {Golestanian}},\ }\href {\doibase 10.1103/PhysRevLett.126.158002} {\bibfield
  {journal} {\bibinfo  {journal} {Physical Review Letters}\ }\textbf {\bibinfo
  {volume} {126}},\ \bibinfo {pages} {158002} (\bibinfo {year}
  {2021}{\natexlab{a}})}\BibitemShut {NoStop}%
\bibitem [{\citenamefont {Mahdisoltani}\ and\ \citenamefont
  {Golestanian}(2021{\natexlab{b}})}]{Mahdisoltani2021}%
  \BibitemOpen
  \bibfield  {author} {\bibinfo {author} {\bibfnamefont {S.}~\bibnamefont
  {Mahdisoltani}}\ and\ \bibinfo {author} {\bibfnamefont {R.}~\bibnamefont
  {Golestanian}},\ }\href {\doibase 10.1088/1367-2630/ac0f1a} {\bibfield
  {journal} {\bibinfo  {journal} {New Journal of Physics}\ }\textbf {\bibinfo
  {volume} {23}},\ \bibinfo {pages} {073034} (\bibinfo {year}
  {2021}{\natexlab{b}})}\BibitemShut {NoStop}%
\bibitem [{\citenamefont {D{\'{e}}mery}\ and\ \citenamefont
  {Dean}(2015)}]{Demery2015a}%
  \BibitemOpen
  \bibfield  {author} {\bibinfo {author} {\bibfnamefont {V.}~\bibnamefont
  {D{\'{e}}mery}}\ and\ \bibinfo {author} {\bibfnamefont {D.~S.}\ \bibnamefont
  {Dean}},\ }\href {\doibase 10.1088/1742-5468/2016/02/023106} {\bibfield
  {journal} {\bibinfo  {journal} {J. Stat. Mech.}\ ,\ \bibinfo {pages}
  {023106}} (\bibinfo {year} {2015})}\BibitemShut {NoStop}%
\bibitem [{\citenamefont {Frusawa}(2020)}]{Frusawa2020}%
  \BibitemOpen
  \bibfield  {author} {\bibinfo {author} {\bibfnamefont {H.}~\bibnamefont
  {Frusawa}},\ }\href {\doibase 10.3390/e22010034} {\bibfield  {journal}
  {\bibinfo  {journal} {Entropy}\ }\textbf {\bibinfo {volume} {22}},\ \bibinfo
  {pages} {34} (\bibinfo {year} {2020})}\BibitemShut {NoStop}%
\bibitem [{\citenamefont {Frusawa}(2022)}]{Frusawa2022}%
  \BibitemOpen
  \bibfield  {author} {\bibinfo {author} {\bibfnamefont {H.}~\bibnamefont
  {Frusawa}},\ }\href {\doibase 10.1039/d1sm01811f} {\bibfield  {journal}
  {\bibinfo  {journal} {Soft Matter}\ }\textbf {\bibinfo {volume} {18}},\
  \bibinfo {pages} {4280} (\bibinfo {year} {2022})}\BibitemShut {NoStop}%
\bibitem [{\citenamefont {Avni}\ \emph {et~al.}(2022)\citenamefont {Avni},
  \citenamefont {Adar}, \citenamefont {Andelman},\ and\ \citenamefont
  {Orland}}]{Avni2022}%
  \BibitemOpen
  \bibfield  {author} {\bibinfo {author} {\bibfnamefont {Y.}~\bibnamefont
  {Avni}}, \bibinfo {author} {\bibfnamefont {R.~M.}\ \bibnamefont {Adar}},
  \bibinfo {author} {\bibfnamefont {D.}~\bibnamefont {Andelman}}, \ and\
  \bibinfo {author} {\bibfnamefont {H.}~\bibnamefont {Orland}},\ }\href
  {\doibase 10.1103/PhysRevLett.128.098002} {\bibfield  {journal} {\bibinfo
  {journal} {Physical Review Letters}\ }\textbf {\bibinfo {volume} {128}},\
  \bibinfo {pages} {98002} (\bibinfo {year} {2022})}\BibitemShut {NoStop}%
\bibitem [{\citenamefont {Chaikin}\ and\ \citenamefont {Lubensky}()}]{Chaikin}%
  \BibitemOpen
  \bibfield  {author} {\bibinfo {author} {\bibfnamefont {P.~M.}\ \bibnamefont
  {Chaikin}}\ and\ \bibinfo {author} {\bibfnamefont {T.~C.}\ \bibnamefont
  {Lubensky}},\ }\href@noop {} {\emph {\bibinfo {title} {{Principles of
  condensed matter physics}}}}\ (\bibinfo  {publisher} {Cambridge University
  Press})\BibitemShut {NoStop}%
\bibitem [{\citenamefont {Hohenberg}\ and\ \citenamefont
  {Halperin}(1977)}]{Hohenberg1977}%
  \BibitemOpen
  \bibfield  {author} {\bibinfo {author} {\bibfnamefont {P.~C.}\ \bibnamefont
  {Hohenberg}}\ and\ \bibinfo {author} {\bibfnamefont {B.~I.}\ \bibnamefont
  {Halperin}},\ }\href {\doibase 10.1103/RevModPhys.49.435} {\bibfield
  {journal} {\bibinfo  {journal} {Reviews of Modern Physics}\ }\textbf
  {\bibinfo {volume} {49}},\ \bibinfo {pages} {435} (\bibinfo {year}
  {1977})}\BibitemShut {NoStop}%
\bibitem [{\citenamefont {Gardiner}(1985)}]{Gardiner1985}%
  \BibitemOpen
  \bibfield  {author} {\bibinfo {author} {\bibfnamefont {C.~W.}\ \bibnamefont
  {Gardiner}},\ }\href@noop {} {\emph {\bibinfo {title} {{Handbook of
  Stochastic Methods}}}}\ (\bibinfo  {publisher} {Springer},\ \bibinfo {year}
  {1985})\BibitemShut {NoStop}%
\bibitem [{\citenamefont {Louis}\ \emph
  {et~al.}(2000{\natexlab{b}})\citenamefont {Louis}, \citenamefont {Bolhuis},
  \citenamefont {Hansen},\ and\ \citenamefont {Meijer}}]{Louis2000a}%
  \BibitemOpen
  \bibfield  {author} {\bibinfo {author} {\bibfnamefont {A.~A.}\ \bibnamefont
  {Louis}}, \bibinfo {author} {\bibfnamefont {P.~G.}\ \bibnamefont {Bolhuis}},
  \bibinfo {author} {\bibfnamefont {J.~P.}\ \bibnamefont {Hansen}}, \ and\
  \bibinfo {author} {\bibfnamefont {E.~J.}\ \bibnamefont {Meijer}},\ }\href
  {\doibase 10.1103/PhysRevLett.85.2522} {\bibfield  {journal} {\bibinfo
  {journal} {Physical Review Letters}\ }\textbf {\bibinfo {volume} {85}},\
  \bibinfo {pages} {2522} (\bibinfo {year} {2000}{\natexlab{b}})}\BibitemShut
  {NoStop}%
\bibitem [{\citenamefont {Coslovich}\ and\ \citenamefont
  {Ikeda}(2013)}]{Coslovich2013}%
  \BibitemOpen
  \bibfield  {author} {\bibinfo {author} {\bibfnamefont {D.}~\bibnamefont
  {Coslovich}}\ and\ \bibinfo {author} {\bibfnamefont {A.}~\bibnamefont
  {Ikeda}},\ }\href {\doibase 10.1039/c3sm50368b} {\bibfield  {journal}
  {\bibinfo  {journal} {Soft Matter}\ }\textbf {\bibinfo {volume} {9}},\
  \bibinfo {pages} {6786} (\bibinfo {year} {2013})}\BibitemShut {NoStop}%
\bibitem [{\citenamefont {Krekelberg}\ \emph {et~al.}(2009)\citenamefont
  {Krekelberg}, \citenamefont {Kumar}, \citenamefont {Mittal}, \citenamefont
  {Errington},\ and\ \citenamefont {Truskett}}]{Krekelberg2009}%
  \BibitemOpen
  \bibfield  {author} {\bibinfo {author} {\bibfnamefont {W.~P.}\ \bibnamefont
  {Krekelberg}}, \bibinfo {author} {\bibfnamefont {T.}~\bibnamefont {Kumar}},
  \bibinfo {author} {\bibfnamefont {J.}~\bibnamefont {Mittal}}, \bibinfo
  {author} {\bibfnamefont {J.~R.}\ \bibnamefont {Errington}}, \ and\ \bibinfo
  {author} {\bibfnamefont {T.~M.}\ \bibnamefont {Truskett}},\ }\href {\doibase
  10.1103/PhysRevE.79.031203} {\bibfield  {journal} {\bibinfo  {journal} {Phys.
  Rev. E}\ }\textbf {\bibinfo {volume} {79}},\ \bibinfo {pages} {031203}
  (\bibinfo {year} {2009})}\BibitemShut {NoStop}%
\bibitem [{\citenamefont {Mausbach}\ and\ \citenamefont
  {May}(2006)}]{Mausbach2006}%
  \BibitemOpen
  \bibfield  {author} {\bibinfo {author} {\bibfnamefont {P.}~\bibnamefont
  {Mausbach}}\ and\ \bibinfo {author} {\bibfnamefont {H.~O.}\ \bibnamefont
  {May}},\ }\href {\doibase 10.1016/j.fluid.2006.07.021} {\bibfield  {journal}
  {\bibinfo  {journal} {Fluid Phase Equilibria}\ }\textbf {\bibinfo {volume}
  {249}},\ \bibinfo {pages} {17} (\bibinfo {year} {2006})}\BibitemShut
  {NoStop}%
\bibitem [{\citenamefont {Jacquin}\ and\ \citenamefont
  {Berthier}(2010)}]{Jacquin2010}%
  \BibitemOpen
  \bibfield  {author} {\bibinfo {author} {\bibfnamefont {H.}~\bibnamefont
  {Jacquin}}\ and\ \bibinfo {author} {\bibfnamefont {L.}~\bibnamefont
  {Berthier}},\ }\href {\doibase 10.1039/b926412d} {\bibfield  {journal}
  {\bibinfo  {journal} {Soft Matter}\ }\textbf {\bibinfo {volume} {6}},\
  \bibinfo {pages} {2970} (\bibinfo {year} {2010})}\BibitemShut {NoStop}%
\bibitem [{\citenamefont {Thompson}\ \emph {et~al.}(2022)\citenamefont
  {Thompson}, \citenamefont {Aktulga}, \citenamefont {Berger}, \citenamefont
  {Bolintineanu}, \citenamefont {Brown}, \citenamefont {Crozier}, \citenamefont
  {in~'t Veld}, \citenamefont {Kohlmeyer}, \citenamefont {Moore}, \citenamefont
  {Nguyen}, \citenamefont {Shan}, \citenamefont {Stevens}, \citenamefont
  {Tranchida}, \citenamefont {Trott},\ and\ \citenamefont {Plimpton}}]{LAMMPS}%
  \BibitemOpen
  \bibfield  {author} {\bibinfo {author} {\bibfnamefont {A.~P.}\ \bibnamefont
  {Thompson}}, \bibinfo {author} {\bibfnamefont {H.~M.}\ \bibnamefont
  {Aktulga}}, \bibinfo {author} {\bibfnamefont {R.}~\bibnamefont {Berger}},
  \bibinfo {author} {\bibfnamefont {D.~S.}\ \bibnamefont {Bolintineanu}},
  \bibinfo {author} {\bibfnamefont {W.~M.}\ \bibnamefont {Brown}}, \bibinfo
  {author} {\bibfnamefont {P.~S.}\ \bibnamefont {Crozier}}, \bibinfo {author}
  {\bibfnamefont {P.~J.}\ \bibnamefont {in~'t Veld}}, \bibinfo {author}
  {\bibfnamefont {A.}~\bibnamefont {Kohlmeyer}}, \bibinfo {author}
  {\bibfnamefont {S.~G.}\ \bibnamefont {Moore}}, \bibinfo {author}
  {\bibfnamefont {T.~D.}\ \bibnamefont {Nguyen}}, \bibinfo {author}
  {\bibfnamefont {R.}~\bibnamefont {Shan}}, \bibinfo {author} {\bibfnamefont
  {M.~J.}\ \bibnamefont {Stevens}}, \bibinfo {author} {\bibfnamefont
  {J.}~\bibnamefont {Tranchida}}, \bibinfo {author} {\bibfnamefont
  {C.}~\bibnamefont {Trott}}, \ and\ \bibinfo {author} {\bibfnamefont {S.~J.}\
  \bibnamefont {Plimpton}},\ }\href {\doibase 10.1016/j.cpc.2021.108171}
  {\bibfield  {journal} {\bibinfo  {journal} {Comp. Phys. Comm.}\ }\textbf
  {\bibinfo {volume} {271}},\ \bibinfo {pages} {108171} (\bibinfo {year}
  {2022})}\BibitemShut {NoStop}%
\end{thebibliography}


\end{document}